%% file: 00_main.tex
\documentclass[times,a4paper,fleqn]{cas-dc}

\usepackage{cas-common}

\usepackage{amsmath,amsfonts}

\usepackage{subcaption} 
\usepackage{booktabs}
\usepackage{multirow}

\usepackage{soul}
\usepackage{float} 
\usepackage{url}
\usepackage{enumitem}
\usepackage{graphicx} 
\usepackage{lipsum} 
\usepackage{natbib}
\usepackage{hyperref}
\usepackage{natbib}
\usepackage{amssymb}
\usepackage[acronym]{glossaries}
\usepackage{cleveref}

\newsavebox{\twosubbox} 

\begin{document}
\let\WriteBookmarks\relax
\def\floatpagepagefraction{1}
\def\textpagefraction{.001}

\shorttitle{Enhancing Reconstruction-Based OOD Detection in Brain MRI with Model and Metric Ensembles}

\shortauthors{E.M.C. Huijben et al.}

\title [mode = title]{Enhancing Reconstruction-Based Out-of-Distribution Detection in Brain MRI with Model and Metric Ensembles}

\author[1]{Evi M. C. Huijben}[orcid=0000-0002-7406-9241]

\cormark[1]
\ead{e.m.c.huijben@tue.nl}

\credit{Conceptualization, Data curation, Formal analysis, Investigation, Methodology, Software, Validation, Visualization, Writing – original draft, Writing – review and editing}

\address[1]{Eindhoven University of Technology, Department of Biomedical Engineering, Medical Image Analysis Group, Eindhoven, The Netherlands}

\author[1,2]{Sina Amirrajab}[orcid=0000-0001-8226-7777]
\address[2]{The D-Lab, Department of Precision Medicine, GROW - Research Institute for Oncology and Reproduction, Maastricht University, Maastricht, the Netherlands}
\credit{Conceptualization, Methodology, Software, Supervision, Writing – review and editing}

\author[1]{Josien P. W. Pluim}
\credit{Conceptualization, Funding acquisition, Project administration, Resources, Supervision, Writing – review and editing}


\cortext[cor1]{Corresponding author}

\begin{abstract}
Out-of-distribution (OOD) detection is crucial for safely deploying automated medical image analysis systems, as abnormal patterns in images could hamper their performance. However, OOD detection in medical imaging remains an open challenge, and we address three gaps: the underexplored potential of a simple OOD detection model, the lack of optimization of deep learning strategies specifically for OOD detection, and the selection of appropriate reconstruction metrics. In this study, we investigated the effectiveness of a reconstruction-based autoencoder for unsupervised detection of synthetic artifacts in brain MRI. We evaluated the general reconstruction capability of the model, analyzed the impact of the selected training epoch and reconstruction metrics, assessed the potential of model and/or metric ensembles, and tested the model on a dataset containing a diverse range of artifacts. Among the metrics assessed, the contrast component of SSIM and LPIPS consistently outperformed others in detecting homogeneous circular anomalies. By combining two well-converged models and using LPIPS and contrast as reconstruction metrics, we achieved a pixel-level area under the Precision-Recall curve of 0.66. Furthermore, with the more realistic OOD dataset, we observed that the detection performance varied between artifact types; local artifacts were more difficult to detect, while global artifacts showed better detection results. These findings underscore the importance of carefully selecting metrics and model configurations, and highlight the need for tailored approaches, as standard deep learning approaches do not always align with the unique needs of OOD detection.

\end{abstract}


\begin{keywords}
Out-of-distribution detection \sep Anomaly detection \sep Deep learning \sep Autoencoder \sep Brain MRI
\end{keywords}

\maketitle

\input{01_Introduction}
\input{02_Related_work}

\input{03_Data}

\input{04_Methods}
\input{05_Experiments}

\input{06_Results}

\input{07_Discussion}

\printcredits

\section*{Acknowledgments}
This work was funded by the Irène Curie Fellowship program and the Biomedical Engineering department of Eindhoven University of Technology.

\section*{Ethical standards}
This research study was conducted retrospectively using human subject data made available in open access. Details regarding ethical approval are stated in the license attached with the open-access datasets.

\section*{Declaration of competing interest}
We declare we do not have conflicts of interest.

\section*{Declaration of generative AI}
During the preparation of this work the authors used DeepL Write and ChatGPT in order to enhance the writing structure and refine grammar. After using these tools, the authors reviewed and edited the content as needed and take full responsibility for the content of the publication.

\bibliographystyle{model2-names}
\bibliography{references}

\newpage
\onecolumn
\noindent{\LARGE Supplementary material}\\
\appendix
\renewcommand{\thefigure}{\thesection.\arabic{figure}}
\setcounter{figure}{0}
\input{09_Appendix}

\end{document}

%% file: 01_Introduction.tex
\section{Introduction}
\label{OODsec:introduction}

Medical imaging plays a critical role in modern healthcare by providing essential information for diagnosis, treatment planning, and disease monitoring. The rapid advancement of imaging technologies and the significant increase in the volume of medical imaging data have led to a growing demand for automated medical image analysis systems. However, these systems often face limitations in robustness and reliability because their performance is highly dependent on the underlying training data distribution \citep{zimmerer2022mood}. Out-of-distribution (OOD) detection and anomaly detection techniques are essential in this context, as they enable the identification of abnormal patterns in medical images that could significantly hamper the performance of AI models \citep{yang2021generalized}. OOD detection refers to the identification of data that differs from the distribution on which a model was trained, while anomaly detection focuses on the detection of localized abnormalities, such as tumors or lesions in medical images. Although the applications of the detection principles are different, both methods share the same fundamental goal: to identify inputs that do not belong to the training data distribution.

OOD detection in medical imaging is closely tied to the availability of benchmark datasets, which defines what is considered in-distribution (ID) and OOD data. Although this is a significant challenge, our work focuses on addressing three gaps. First, there is a growing trend toward increasingly complex models, yet the potential of simpler models remains underexplored. Second, many current deep learning methods are designed for tasks where training and test data are drawn from the same distribution, but these methods are rarely optimized for OOD detection. Third, the selection of appropriate reconstruction metrics is rarely prioritized, yet it is crucial for improving OOD detection performance.

In order to develop and benchmark methods for OOD detection in medical images, the Medical Out-of-Distribution (MOOD) challenge\footnote{\url{http://medicalood.dkfz.de/web/}} \citep{zimmerer2022mood} was introduced, first organized in conjunction with MICCAI 2020. The MOOD challenge includes sample-level and pixel-level anomaly detection for brain magnetic resonance imaging (MRI) and abdominal computed tomography (CT) datasets. In this context, sample-level anomaly detection refers to identifying whether an entire image (or sample) belongs to the training distribution, while pixel-level anomaly detection focuses on identifying local abnormal regions within the image, such as lesions or tumors.

In this work, we investigated whether a reconstruction-based autoencoder (AE) could effectively detect anomalies and artifacts in brain MRI in an unsupervised manner. We first evaluated the trade-off between latent space size and reconstruction capability to determine optimal model configurations. We then examined whether the absolute reconstruction error is sufficient for anomaly detection or whether alternative metrics, such as structural similarity index measure (SSIM), perceptual metrics, or including ensembles of different metrics, provide better performance. These experiments were initially conducted on an OOD dataset with homogeneous circular anomalies to select the optimal configuration. We then extended our analysis to an extended brain MRI dataset containing three types of local artifacts and seven types of global artifacts. Through these experiments, we analyzed deep learning design decisions and proposed refinements to improve OOD detection in medical imaging. In summary, our contributions are as follows:
\begin{itemize}
    \item We investigate the optimal latent space size of an AE for OOD detection.
    \item We critically analyzed the choice of training epoch and metrics and explored alternatives to the standard choices.
    \item We explore the use of model and/or metric ensembles to improve OOD detection.
    \item  We generate two datasets containing local and/or global artifacts specific to brain MRI for performance evaluation.
    \item  We conducted an in-depth analysis of artifact severity and its effect on OOD detection performance.
\end{itemize}

To support reproducibility and encourage further research, the code for generating the synthetic OOD datasets and for training and testing the model is publicly available at \url{https://github.com/evihuijben/ood_detection_mri}.

%% file: 02_Related_work.tex
\section{Related work}
\label{OODsec:related work}
OOD detection methods can be categorized into recon\-struction-based, restoration-based, and generative-based approaches \citep{baur2021autoencoders, graham2023denoising}. Recon\-struction-based methods work by training encoder-decoder models to accurately reconstruct ID inputs. OOD detection is then performed by identifying samples that exhibit poor reconstruction performance. Restoration-based methods build upon this by manipulating the latent space of the encoded images to actively guide the output toward ID samples, effectively "healing" anomalies in the reconstructed images. Generative methods use a probability distribution model of ID data and use the likelihood to identify OOD samples. These methods can be implemented in an unsupervised setting, relying solely on healthy (i.e., ID) images for training, or in a self-supervised setting, where synthetic anomalies are introduced during training to improve detection. In \autoref{subsec:rw_models}, we discuss different models that are used in medical image OOD detection, and in \autoref{subsec:rw_metrics}, we discuss the various metrics proposed to obtain an anomaly score, either at pixel-level or at sample-level.

\subsection{OOD detection models in medical imaging}
\label{subsec:rw_models}
AEs are widely used for unsupervised anomaly detection (UAD), relying on the calculation of the reconstruction error of only ID samples during training to identify anomalies at test-time \citep{baur2021autoencoders, zimmerer2019unsupervised}. Despite their popularity for UAD in medical imaging, challenges remain in accurately learning the healthy distribution and developing a robust model that is not tailored to detect specific types of anomalies \citep{bercea2023generalizing}. To address these challenges, efforts have been made to overcome them using a scale-space AE \citep{baur2020scale}, memory-augmented AE \citep{gong2019memorizing}, reversed AE \citep{bercea2023generalizing}. \citet{cai2024rethinking} further contributed by using information theory to guide the design of optimal AEs by limiting the size of the latent space and aligning the entropy of the latent space with the entropy of the normal data. Furthermore, encoder-decoder structures are also a popular choice for self-supervised anomaly detection where the model is trained on images with synthetically introduced anomalies \citep{tan2021detecting, baugh2022nnood, kascenas2022denoising, baugh2023many,p2023confidence}.

\citet{baur2021autoencoders} found that variational AEs (VAEs) \citep{kingma2013auto}, generally performed better in UAD than traditional AEs. Although VAEs often resulted in blurry reconstructions, they showed larger differences between the reconstruction errors of normal and anomalous pixels. In particular, image restoration using a VAE with normative prior \citep{you2019unsupervised} outperformed reconstruction-based VAEs \citep{baur2021autoencoders}. \citet{wang2020image} proposed to use vector quantized (VQ)-VAEs, which employ a discrete latent space, to further improve the image restoration process by restoring OOD latent representations. Furthermore, generative-based OOD detection using VAEs has also been proposed, with lower likelihoods indicating OOD samples \citep{an2015variational}. \citet{zimmerer2019unsupervised} extended this approach by incorporating reconstruction probability and a term derived from the Kullback–Leibler (KL) divergence.

Generative adversarial networks (GANs), such as f-AnoGAN \citep{schlegl2019f} and GANomaly \citep{akcay2019ganomaly}, have shown strong performance in medical imaging OOD detection by leveraging reconstruction-based anomaly scoring \citep{baur2021autoencoders,van2021anomaly}. Transformer-based models have also achieved state-of-the-art results in this domain \citep{graham2022transformer, pinaya2022unsupervised, lu2024anomaly}, although they face scalability and efficiency challenges.

\sloppy
Recently, denoising probabilistic diffusion models (DDPMs) \citep{ho2020denoising}, introduced as a powerful category of generative models, have found their way into unsupervised anomaly detection in medical images using a partial diffusion process \citep{wyatt2022anoddpm, graham2023denoising, wolleb2022diffusion,huijben2023histogram}. DDPMs face scalability issues, and that is why latent diffusion models (LDMs) \citep{rombach2022high} have been introduced for OOD detection \citep{pinaya2022brain,graham2023unsupervised}. Instead of the standard Gaussian noise, alternative noises such as simplex noise \citep{wyatt2022anoddpm}, coarse noise \citep{kascenas2023role}, and Bernoulli noise for a binary LDM \citep{wolleb2024binary} have been proposed for OOD detection. \citet{naval2024disyre} and \citet{sanchez2022healthy} went a step further and used a counterfactual diffusion process by gradually introducing synthetic anomalies for weakly supervised anomaly detection. 

A drawback of using DDPMs for OOD detection is that healthy tissue is often altered in the reverse diffusion process. To overcome this and preserve the integrity of healthy tissue, various masking and stitching mechanisms have been explored: patched DDPM \citep{behrendt2024patched}, masked DDPM \citep{iqbal2023unsupervised}, AutoDDPM \citep{bercea2023mask}, masking in the latent space of an LDM \citep{pinaya2022fast, wolleb2024binary} and thermal harmonization for optimal restoration \citep{bercea2024diffusion}.

\subsection{Anomaly scoring}
\label{subsec:rw_metrics}

Reconstruction-based approaches typically quantify OOD detection using various reconstruction metrics, such as the reconstruction error (absolute or squared error), the SSIM \citep{wang2004image}, and the learned perceptual image patch similarity (LPIPS) \citep{zhang2018unreasonable}. These metrics are used directly in many of the studies discussed in \autoref{subsec:rw_models}. However, \citet{meissen2022pitfalls} highlighted the significant limitations of using reconstruction error for grayscale images as it depends on the underlying intensities and suggested applying post-processing techniques to these reconstruction error maps. This has led to the adoption of techniques such as median filtering, normalization, connected component analysis, morphological operations (e.g., erosion, dilation, closing), and thresholding of the final anomaly prediction map \citep{baur2021autoencoders, kascenas2022denoising, huijben2023histogram,bercea2023mask}. In addition, \citet{bercea2023generalizing} used adaptive histogram equalization on the input and reconstructed images prior to metric computation. A common strategy to reduce false positives is to apply a body contour mask \citep{baur2021autoencoders, huijben2023histogram, kascenas2023role}. Standardizing results based on statistics of an ID dataset (e.g., the training or validation set) is also a common approach \citep{solal2024leveraging}. \citet{bercea2023evaluating} take this a step further and proposed three normative learning-based metrics: 1) the restoration quality index (RQI) based on LPIPS, 2) the anomaly to healthy index (AHI) based on the Fréchet inception distance (FID) \citep{heusel2017gans}, and 3) the conservation and anomaly correction index (CACI) based on the SSIM.

Some model architectures support combining metric maps over multiple resolutions \citep{baur2020scale} or modalities \citep{kascenas2022denoising, kascenas2023role}. \citet{lu2024anomaly} used multi-stage feature cosine similarity to generate anomaly score maps. Other methods incorporate information from multiple latent vectors \citep{akcay2019ganomaly}, intermediate latent predictions \citep{pinaya2022fast}, or discriminator features \citep{zhang2022unsupervised}. Focusing on SSIM, several ensemble approaches have become popular in OOD detection, including multi-scale SSIM \citep{tian2023unsupervised}, multi-scale feature space SSIM \citep{meissen2022unsupervised}, and SSIM ensembles using multiple window sizes \citep{behrendt2024diffusion}.

Diffusion models inherently have multiple reconstructions due to their iterative nature, allowing a metric to be computed for all reconstructions \citep{graham2023unsupervised, wolleb2024binary, naval2024ensembled,bercea2024diffusion}. Taking advantage of this property, \citet{behrendt2024leveraging} proposed to leverage the Mahalanobis distance \citep{mahalanobis1936generalised} across multiple reconstructions.

%% file: 03_Data.tex
\section{Data}
\label{sec:data}

We used the publicly available brain MRI dataset from the MOOD challenge\footnote{\url{https://www.synapse.org/Synapse:syn21343101/wiki/599515}}. This dataset contains a brain MRI scan for 800 healthy individuals. The resolution of each scan is $256\times256\times256$ voxels, and we linearly rescaled the intensities of the 3D volumes to $[0,1]$. 

\subsection{Training and validation sets}
\begin{figure*}
    \centering
    \includegraphics[width=\linewidth]{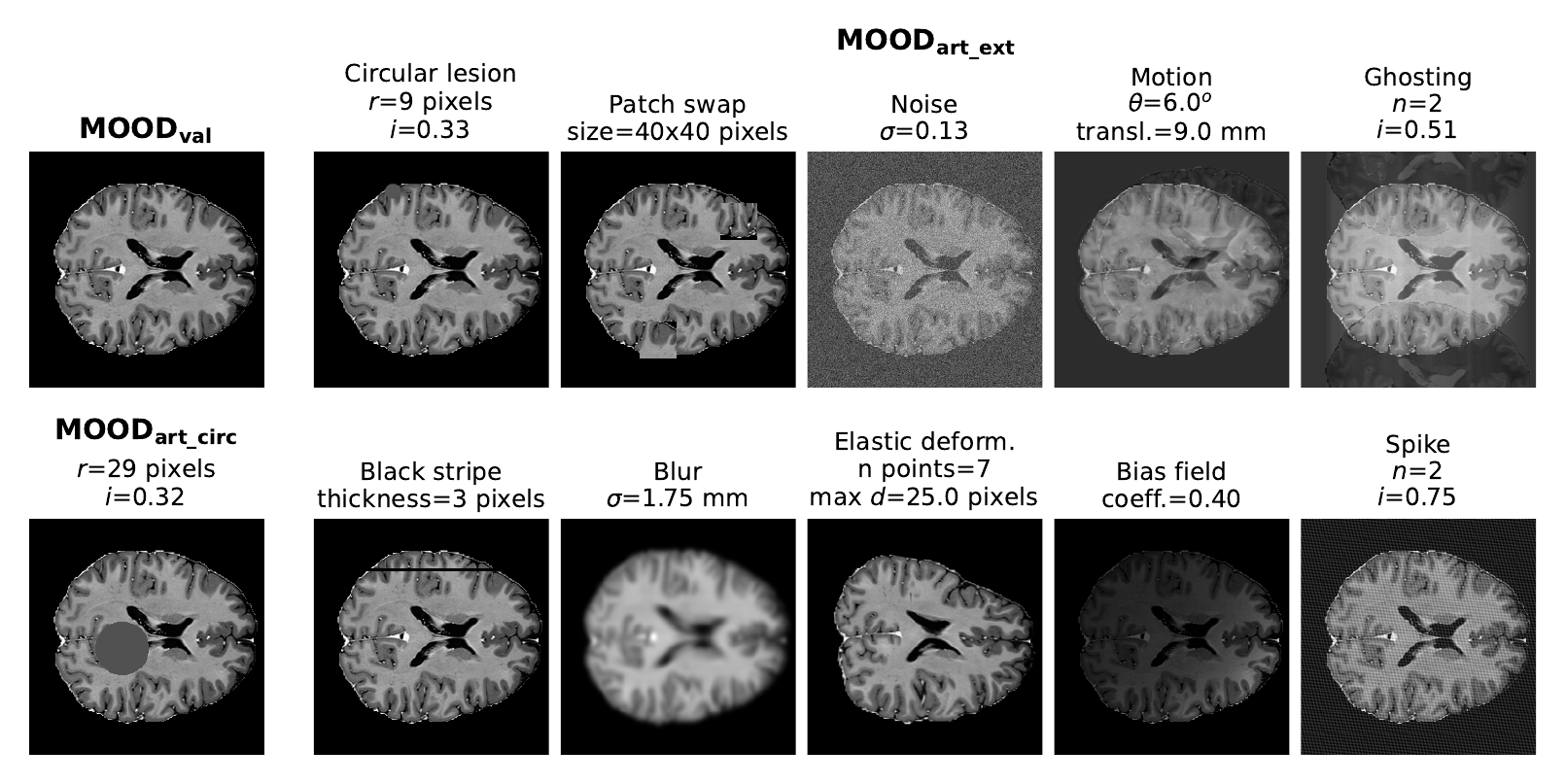}
    \caption{Example image from the MOOD\textsubscript{val} dataset (upper left), its counterpart from MOOD\textsubscript{art\_circ} (lower left), and the counterparts from MOOD\textsubscript{art\_ext} (second to last column). Additionally, the parameter values used to generate the artifacts are presented in accordance with the explanation provided in \autoref{OODtab:MOOD_aug details}.}
    \label{fig:dataset}
\end{figure*}

We randomly divided the MOOD dataset into an ID training set (720 cases) and ID validation set (80 cases). For both the training and validation sets, we included the middle 20 axial slices to create a homogeneous dataset for a 2D implementation and evaluation. We call these sets of axial slices MOOD\textsubscript{train} and MOOD\textsubscript{val}, respectively. An example of a slice from MOOD\textsubscript{val} is shown in the upper left corner of \autoref{fig:dataset}.

\subsection{Validation set with circular artifacts}
In addition, we created an OOD validation set with synthetic circular artifacts called MOOD\textsubscript{art\_circ}. This set is based on MOOD\textsubscript{val} to facilitate direct comparisons between the images in the two sets and to allow parameter tuning.  To generate MOOD\textsubscript{art\_circ}, a homogeneous circle was added to the 1600 axial brain slices in MOOD\textsubscript{val}.
This was done by selecting a non-zero voxel as the center and then randomly selecting a radius $r \sim U([20, 40])$ and an intensity $i \sim U(0,1)$. The added circle inherently provides a ground truth segmentation label for pixel-level evaluation. An example of a slice from MOOD\textsubscript{art\_circ} is shown in the lower left corner of \autoref{fig:dataset}.

\subsection{Extended synthetic artifact dataset}
An extended set with more types of synthetic artifacts, called MOOD\textsubscript{art\_ext}, was created based on MOOD\textsubscript{val} to evaluate the performance of the model on more realistic artifacts. Specifically, we introduced three types of local artifacts: circular lesion (similar to MOOD\textsubscript{art\_circ} but smaller circles), black stripe, and patch swap. In addition, we introduced seven types of global artifacts: blurring, noise, elastic deformation, patient motion, MRI bias field, MRI ghosting artifact, and MRI spikes (also known as Herringbone artifact). For each artifact type, artifacts were introduced for the entire validation set, resulting in ten new validation sets of 1600 slices to allow direct comparison with MOOD\textsubscript{val}. All artifacts were systematically introduced and increased in severity in $N$ steps. The parameters for the creation of these artifacts are summarized in \autoref{OODtab:MOOD_aug details} and an example for each artifact is shown in \autoref{fig:dataset} (second to last columns). Note that due to the range of artifact parameter values, some resulting slices may not be distinctly OOD as the severity of the artifact gradually increases. For example, a slice with subtle elastic deformation could be a naturally occurring variation rather than being OOD.

A detailed description on the creation of each artifact type can be found in Appendix A and our code is publicly available at \url{https://github.com/evihuijben/ood_detection_mri}.

\begin{table}
\footnotesize
\centering
\renewcommand{\arraystretch}{1.2}  
\caption{Details of artificially introduced anomalies and artifacts in the MOOD\textsubscript{art\_ext} dataset. The value for each parameter was linearly increased from min to max in $N$ steps.}
\begin{tabular}{llll}
\toprule
\textbf{Local artifact} & \textbf{Parameter [unit]} & \textbf{[min, max]} & \textbf{N} \\

\midrule
\multirow{2}{*}{Circular lesion} & radius (\(r\)) [pixels] & [3, 30] & 10 \\
 & intensity (\(i\)) [-] & [0, 1] &  10\\

\midrule
\multirow{1}{*}{Black stripe} & thickness [pixels] & [1, 5] & 5 \\

\midrule
\multirow{1}{*}{Patch swap} & width/height [pixels] & [30, 70] & 10 \\

\bottomrule
\toprule
\textbf{Global artifact} & \textbf{Parameter [unit]} & \textbf{[min, max]} & \textbf{N} \\
\midrule
\multirow{1}{*}{Blur} & standard dev. (\(\sigma\)) [mm] & [0.25, 2.5] & 10 \\

\midrule
\multirow{1}{*}{Noise} & standard dev. (\(\sigma\))  [-] & [0.01, 0.37] & 10 \\

\midrule
\multirow{2}{*}{Elastic deform.} & n control points / axis [-] & [5, 8] & 4 \\
 & max displ. (\(d\)) [pixels] &  [7.5, 30] & 10 \\

 \midrule
\multirow{2}{*}{Motion} & rotation [$^o$] & [1, 10] & 10 \\
 & translation [mm] & [1.5, 15] & 10\\

\midrule
\multirow{1}{*}{Bias field} & coefficients [-] & [0.05, 0.5] & 10 \\

\midrule
\multirow{2}{*}{Ghosting} & Number of ghosts (\(n\)) [-] & [1, 2] & 2 \\
 & intensity (\(i\)) [-]& [0.4, 0.6] & 10 \\

\midrule
\multirow{2}{*}{Spike} & number of spikes (\(n\)) [-] & [1, 2] & 2\\
 & intensity (\(i\)) [-] &  [0.25, 2.5] &  10\\

\bottomrule
\end{tabular}

\label{OODtab:MOOD_aug details}
\end{table}

%% file: 04_Methods.tex
\section{Methods}
\label{OODsec:methods}
\subsection{OOD detection model}
\begin{figure*}
    \centering
    \includegraphics[width=\linewidth]{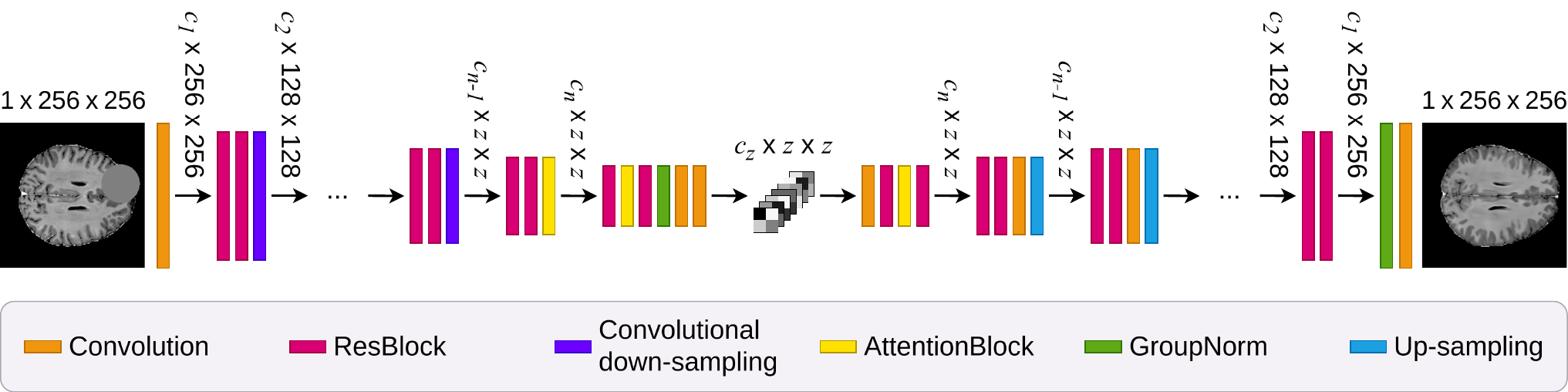}
    \caption[The neural network architecture used in this work. The exact details of the layers can be found in the original implementation of AutoencoderKL by MONAI generative]{The neural network architecture used in this work. The exact details of the layers can be found in the original implementation of AutoencoderKL by MONAI generative\footnotemark.}
    \label{fig:network}
\end{figure*}

In this work, we use an AE as implemented in the MONAI generative models library \citep{pinaya2023generative} for the task of OOD detection. Specifically, the architecture of the KL-regularized latent space AE as described in \citet{rombach2022high} and \citet{pinaya2022brain} was used. However, since our task is not to generate samples, but only to reconstruct, we have removed the latent space sampling and the KL divergence. The network architecture consists of a convolutional cascade using residual blocks \citep{he2016deep} followed by down- or up-sampling, as illustrated in \autoref{fig:network}. Furthermore, the model includes a PatchGAN discriminator \citep{wang2018high} to allow for adversarial training to improve the image reconstruction.

\setcounter{footnote}{1} 

We experimented with four versions of the AE with different latent space sizes: $2\times2$, $4\times4$, $8\times8$, and $16\times16$. Each model achieves its respective latent size by modifying the model depth to include $n$ blocks of the double residual blocks, followed by the down- or up-sampling layer. Note that the last block does not use a down- or up-sampling layer. The number of channels used for the feature representations is described in \autoref{tab:network} and the number of latent channels is set to $c_z=32$ unless stated otherwise.

\begin{table}[]
\footnotesize
    \centering
    \begin{tabular}{lllllllll}
    \toprule
    \textbf{$z\times z$} & \textbf{$c_1$}& \textbf{$c_2$} & \textbf{$c_3$} & \textbf{$c_4$} & \textbf{$c_5$} & \textbf{$c_6$} & \textbf{$c_7$}& $c_8$ \\
    \midrule
    $2\times2$ & 128 & 128 & 256 & 256 & 512 &512 & 512 & 512 \\
    $4\times4$ & 128  & 128 & 256& 256 & 512 & 512 &512 \\
    $8\times8$ & 128  & 128 & 256& 256 & 512 & 512 \\
    $16\times16$ & 128  & 128 & 256& 256 & 512  \\
    \bottomrule
    \end{tabular}
    \caption{Channels used for the intermediate features in the encoder and decoder for the models with different latent space sizes ($z\times z$).}
    \label{tab:network}
\end{table}

The models were trained using multiple loss components measuring the difference between the input image $x$ and the reconstruction $\hat{x}$:

\textbf{Reconstruction Loss ($\mathcal{L}_{\text{rec}}$)} measures the fidelity of image reconstruction, calculated as
\begin{equation}
\mathcal{L}_{\text{rec}}(x, \hat{x}) = \|x - \hat{x}\|_1.
\end{equation}
    
\textbf{Perceptual Loss ($\mathcal{L}_{\text{perceptual}}$)} \citep{zhang2018unreasonable} captures the perceptual similarity between $x$ and $\hat{x}$ by computing the difference between features $\delta (x)$ and $\delta (\hat{x})$ obtained from the pre-trained VGG-16 \citep{simonyan2014very}:
\begin{equation}
    \mathcal{L}_{\text{perceptual}} (x, \hat{x}) = \left| \delta(x) - \delta(\hat{x})\right\|_2^2.
\end{equation}
    
\textbf{Adversarial Loss ($\mathcal{L}_{\text{adv}}$)} enhances the model's ability to generate realistic images. We employed the least squares adversarial loss based on a discriminator $D$ as follows:
\begin{equation}
\mathcal{L}_{\text{adv}}(\hat{x}) = \left(D(\hat{x}) - 1 \right)^2 .
\end{equation}

\textbf{Total generator loss ($\mathcal{L}_{\text{gen}}$)} is a weighted combination of the individual losses, calculated as
\begin{equation}
\mathcal{L}_{\text{gen}} = \mathcal{L}_{\text{recons}}  + \alpha \mathcal{L}_{\text{perceptual}} + \beta \mathcal{L}_{\text{adv}},
\end{equation}
where $\alpha$ and $\beta$ are set to 2e-3 and 5e-3.

\textbf{Discriminator loss ($\mathcal{L}_{\text{discr}}$)} is computed as the average of the fake and real adversarial losses:
\begin{equation}
\mathcal{L}_{\text{discr}} = \beta \frac{1}{2} \left( \left(D(\hat{x}) -  0\right)^2 + \left(D(x) - 1\right)^2 \right).
\end{equation}

\footnotetext{\url{https://github.com/Project-MONAI/GenerativeModels}}

\subsection{Implementation details}
Each model was trained for 100 epochs with a batch size of 4 using Adam optimizer. The learning rates for the generator and discriminator were set to 1e-4 and 5e-4, respectively. The optimal epoch, i.e. the model with the lowest $\mathcal{L}_{\text{gen}}$ for MOOD\textsubscript{val}, was selected for inference. All models were implemented in Pytorch 2.2.0 with CUDA version 12.1 and trained on an NVIDIA GeForce RTX 2080-Ti 11GB GPU or an NVIDIA TITAN XP 12GB GPU.

\subsection{Reconstruction metrics}
\label{subsec:recon_metrics}
Six different reconstruction metrics were used to assess the reconstruction quality and identify anomalies. Given an input image \(x\) and its reconstructed image \(\hat{x}\), an error map of the same size as \(x\) and \(\hat{x}\) was calculated for each metric. 

\textbf{Absolute error ($E_{abs}$)} was calculated to measure the absolute difference between corresponding pixels in $x$ and $\hat{x}$, defined as
\begin{equation}
E_{abs}(x, \hat{x}) = | x - \hat{x} |.
\end{equation}

Structural similarity is based on three components: contrast, luminance, and structure. These components are computed using the mean intensities \( \mu_x \) and \( \mu_{\hat{x}} \), the standard deviations \( \sigma_x \) and \( \sigma_{\hat{x}} \), and the covariance \( \sigma_{x\hat{x}} \) between the input image \( x \) and the reconstruction \( \hat{x} \). Small constants \( C_1=0.0001 \), \( C_2=0.0009 \), and \( C_3=C_2/2 \) are included for numerical stability. Practically, SSIM is computed over local windows of \( 7 \times 7 \) pixels to capture local similarity. Additionally, negative values in \( \sigma_x \), \( \sigma_{\hat{x}} \), and \( \sigma_{x\hat{x}} \) were clipped to zero to ensure that SSIM scores remain within the [0, 1] range. The three components are evaluated separately and combined as SSIM and are defined as follows:

\textbf{Contrast (\(c\))} assesses the contrast between the two images, accounting for variations in pixel intensity, defined as
\begin{equation}
c(x, \hat{x}) = \frac{2 \sigma_x \sigma_{\hat{x}} + C_2}{\sigma_x^2 + \sigma_{\hat{x}}^2 + C_2}.
\end{equation}

\textbf{Luminance (\(l\))} measures the brightness of the images, capturing the mean intensity of the pixels in both images, defined as:
\begin{equation}
l(x, \hat{x}) = \frac{2 \mu_x \mu_{\hat{x}} + C_1}{\mu_x^2 + \mu_{\hat{x}}^2 + C_1}.
\end{equation}

\textbf{Structure (\(s\))} captures the structural information between the two images, reflecting how pixel patterns correlate spatially. It is defined as
\begin{equation}
s(x, \hat{x}) = \frac{\sigma_{x\hat{x}} + C_3}{\sigma_x \sigma_{\hat{x}} + C_3}.
\end{equation}

\textbf{Structural similarity index measure (SSIM)} combines these three components to provide an overall measure of similarity between \( x \) and \( \hat{x} \), calculated as 
\begin{equation}
\text{SSIM}(x, \hat{x}) = c(x,\hat{x})\cdot l(x, \hat{x})\cdot s(x,\hat{x})
\end{equation}

\textbf{Learned perceptual image patch similarity (LPIPS)} \citep{zhang2018unreasonable} captures perceptual similarity between \( x \) and \( \hat{x} \) by computing feature-level differences at each spatial location, defined as
\begin{equation}
\text{LPIPS}(x, \hat{x}) = \left\| \phi(x) - \phi(\hat{x}) \right\|_2^2,
\end{equation}
where \( \phi(.)\) represents the features extracted from a pre-trained AlexNet.

\subsection{Quantification of prediction maps}
In order to quantitatively evaluate the overall performance of the reconstruction metrics presented in \autoref{subsec:recon_metrics}, the mean of the respective reconstruction map was computed to obtain one prediction score for each slice.

To facilitate combinations of reconstruction metrics, the reconstruction error maps must be in the range of 0 (indicating a negative prediction) to 1 (indicating a positive prediction). To ensure that all metrics were consistent with this range, we inverted the SSIM and its components, yielding \(1 - \text{SSIM}(x, \hat{x})\), \(1 - c(x, \hat{x})\), \(1 - l(x, \hat{x})\), and \(1 - s(x, \hat{x})\). Then, we averaged multiple (inverted) reconstruction maps to combine metrics, resulting in a final anomaly prediction map \(\mathcal{A}\). 
When evaluating performance per slice, we again took the average of the prediction map \(\mathcal{A}\) to obtain one score per slice, represented by \(\bar{\mathcal{A}}\).

In cases where local anomalies were present and a ground truth mask was available, the prediction task could be seen as a pixel-level classification problem. This approach enabled the generation of a Precision--Recall curve for any (inverted) prediction map with values ranging from 0 to 1. Finally, we quantified the overall performance using the area under the Precision--Recall curve (AUPRC) as suggested by \citet{baur2021autoencoders}. We calculated the precision and recall at thresholds evenly spaced between 0 and 1, with a step size of 0.02.

\subsection{Statistical testing}

Statistical tests were used to assess the statistical significance of the observed differences in mean prediction maps between the data sets.

Specifically, when comparing the results of a validation set with those of the training set, we applied a one-sided Mann-Whitney U test \citep{mann1947test} with a significance level of \(\alpha=0.01\). This non-parametric test, suitable for comparing two independent samples that may not follow a normal distribution, was conducted to assess whether the validation set's performance is significantly greater than that of the training set.

When comparing the performance of one type of artifact in MOOD\textsubscript{art\_ext} with the performance of MOOD\textsubscript{val} dataset, we used a one-sided Wilcoxon signed-rank test \citep{wilcoxon1945} with a significance level of \(\alpha=0.01\). This test assesses whether the performance of the extended artifact dataset is significantly higher than that of the ID validation set, providing insight into pairwise performance differences between the two sets.

%% file: 05_Experiments.tex
\section{Experiments}
\label{OODsec:experiments}

We conducted a set of experiments to 1) evaluate general reconstruction capability of the model, 2) analyze the effect training epoch and overfitting, 3) identify the most effective reconstruction metric, 4) assess the feasibility of combining models and/or metrics, and 5) evaluate the effectiveness of our OOD detection model on a more realistic OOD dataset.

\subsection{Reconstruction capability}
We analyzed the trade-off between a smaller latent space, which limits the model’s ability to optimally reconstruct ID data, and a larger latent space, which enables the reconstruction of OOD data. Specifically, we evaluate the performance of models with latent space sizes of $2\times2$, $4\times4$, $8\times8$, and $16\times16$. The number of channels in the latent space ($c_z$) is set to 32, 64, or 128. We assessed the reconstruction capability on the MOOD\textsubscript{train}, MOOD\textsubscript{val}, and MOOD\textsubscript{art\_circ} sets using the mean $E_{abs}$ (MAE) per slice. Statistical significance is calculated between MOOD\textsubscript{val} and MOOD\textsubscript{train}, as well as between MOOD\textsubscript{art\_circ} and MOOD\textsubscript{train}.

\subsection{Optimal versus final epoch}
To investigate the effect of OOD detection, we analyzed how different training epochs impact the model's ability to generalize. Overfitting may lead the model to memorize training examples, potentially hindering its performance in detecting OOD data, as we aim for the model to learn the overall training distribution rather than specific instances. However, OOD detection is different from traditional deep learning models, where all sets are assumed to be from the same distribution. Therefore, we examined the effect of selecting the optimal and final epoch for the $2\times2$ and $8\times8$ models. Again, we analyzed the MAE of the MOOD\textsubscript{train}, MOOD\textsubscript{val}, and MOOD\textsubscript{art\_circ} datasets and calculated the statistical significance between MOOD\textsubscript{val} and MOOD\textsubscript{train}, and between MOOD\textsubscript{art\_circ} and MOOD\textsubscript{train}.

\subsection{Choice of metric}
The MAE is a limited measure of reconstruction quality because it only captures local intensity changes, in a pixel-by-pixel manner, and does not include perceptual, structural, or contrast components. Therefore, in this experiment, we evaluated the usefulness of different reconstruction metrics. The metrics used were all the reconstruction metrics described in \autoref{subsec:recon_metrics}: absolute error, luminance, contrast, structure, SSIM, and LPIPS. Specifically, we visualize the results of all metrics, and then quantitatively select the optimal metric by calculating the AUPRC of the (inverted) metric maps. We present the results of the $2\times2$ and $8\times8$ models for both the final and optimal epochs.

\subsection{Ensemble of models and metrics}
The strengths of different models and metrics can be combined to achieve the best performance. For this experiment, we analyzed the Precision--Recall curve for different combinations of models and/or metrics. Specifically, we examined the reconstruction errors in terms of LPIPS and contrast of the final epoch for the $2\times2$ and $8\times8$ models. We calculated the final anomaly prediction map \(\mathcal{A}\) by taking the average of LPIPS and $1-c(x,\hat{x})$ of both models.

\subsection{Extended synthetic artifact dataset}
To determine the effectiveness of the model in translating to a more nuanced and realistic artifact dataset, we tested the model on MOOD\textsubscript{art\_ext}. Specifically, the final prediction map for each slice (\(\mathcal{A}\)) was calculated as the average of the LPIPS and $1-c(x,\hat{x})$ maps for both the $2\times2$ and $8\times8$ models. To obtain a single score per slice, the mean of each final prediction map was taken, which we refer to as \(\bar{\mathcal{A}}\). For each artifact type we investigated whether each MOOD\textsubscript{art\_ext} sample had a significantly higher score \(\bar{\mathcal{A}}\) compared to its MOOD\textsubscript{val} counterpart. Naturally, evaluating the local artifacts (circular lesion, black stripe, and patch swap) at the global level using \(\mathcal{\bar{A}}\), yields results that are substantially influenced by the size of the local anomaly. Therefore, for these local anomalies, we also present the AUPRC values calculated based on the final prediction maps \(\mathcal{A}\).

Furthermore, since the extended artifact dataset was systematically constructed based on carefully selected parameter values for each artifact type, we investigated the correlation between parameter values and OOD detection performance. We quantified these correlations by calculating the Spearman rank correlation coefficient \(\rho\) \citep{spearman} between the parameter value and the final score \(\bar{\mathcal{A}}\). This coefficient takes into account the ordinal relationship between the ranks of individual test case performances, providing robustness to variations in the scale and direction of the values.

%% file: 06_Results.tex
\section{Results}
\label{OODsec:results}

\subsection{Reconstruction capability}
\begin{figure}
    \centering
    \includegraphics[width=\linewidth]{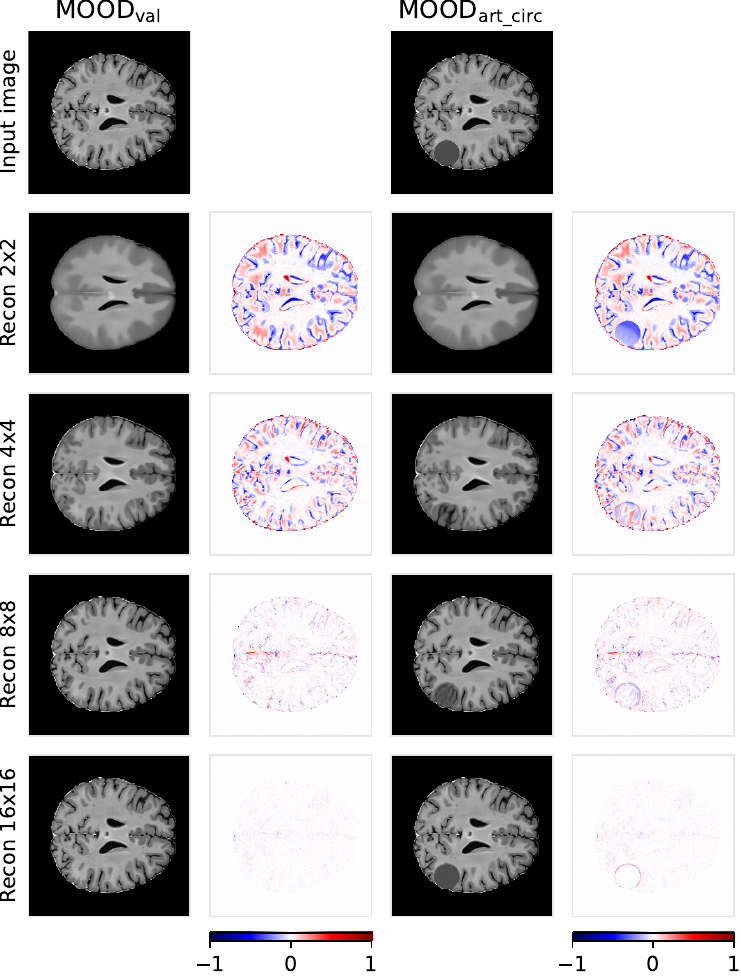}
    \caption{Visual results based on different models with varying depths and fixed number of latent channels of $c_z=32$. The first row shows a random example from MOOD\textsubscript{val} and its counterpart MOOD\textsubscript{art\_circ}. Second to last rows show the reconstructions and the respective errors ($x-\hat{x}$) for models with different latent space sizes.}
    \label{fig:recons_latent_sizes}
\end{figure}
\autoref{fig:recons_latent_sizes} shows the reconstructions of a random MOOD\textsubscript{val} input and its MOOD\textsubscript{art\_circ} counterpart for all the different models with a fixed number of latent channels of $c_z=32$. We observe that the $2\times2$ model seems to reconstruct a rough average brain image. The $4\times4$ model appears to be more successful in preserving the original anatomy of the brain while roughly inpainting the abnormal region. However, because of the still relatively high absolute error in other parts of the brain, the anomalous region is not standing out in the error map. Furthermore, the less deep models with latent spaces of $8\times8$ and $16\times16$ show that they are not able to ``heal'' the anomalous region and therefore not able to detect it based on the reconstruction absolute error.
\begin{figure}
    \centering
    \includegraphics[width=\linewidth]{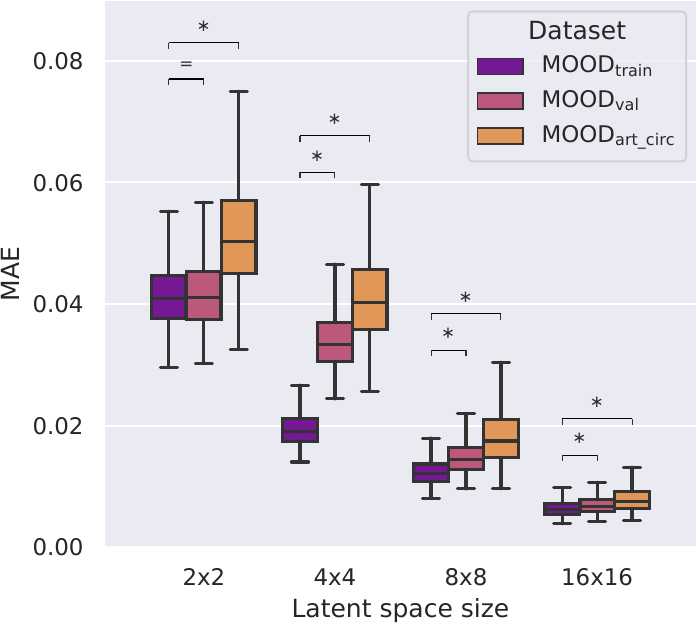}
    \caption{MAE of the reconstructions obtained for different models with varying depths. Statistical significance (\(\alpha=0.01\)) was assessed between MOOD\textsubscript{train} and MOOD\textsubscript{val}, as well as between MOOD\textsubscript{train} and MOOD\textsubscript{art\_circ}. Significant differences are indicated by an asterisk and non-significant differences are indicated by an equal sign. Note that outliers are not shown in the boxplots for visibility purposes.}
    \label{fig:box_latent_sizes}
\end{figure}
\autoref{fig:box_latent_sizes} shows that the overall reconstruction error in terms of MAE per axial slice decreases as the depth of the model increases, and thus the size of the latent space decreases. A well-trained OOD detection model would show a low reconstruction error for both the MOOD\textsubscript{train} and MOOD\textsubscript{val} datasets, and a high reconstruction error for the MOOD\textsubscript{art\_circ} dataset. However, we observe different patterns for the different model depths. The $2\times2$ model achieves the best distinction between MOOD\textsubscript{val} and MOOD\textsubscript{art\_circ}, while maintaining consistency between MOOD\textsubscript{val} and MOOD\textsubscript{train}. On the other hand, the $4\times4$ model is severely overfitted to the training set since it does not generalize well to MOOD\textsubscript{val}. The $8\times8$ and $16\times16$ models show less overfitting, but also do not clearly distinguish between MOOD\textsubscript{val} and MOOD\textsubscript{art\_circ}. Statistical analyses confirm that only for the $2\times2$ model is the reconstruction error for the MOOD\textsubscript{val} set statistically indistinguishable from that of MOOD\textsubscript{train} (\(p=0.035\)). In contrast, all other reconstruction errors for MOOD\textsubscript{val} and MOOD\textsubscript{art\_circ} are significantly higher \((p\sim0.00\)) than for the training set. Based on these observations, the $2\times2$ and $8\times8$ models were used for further experiments.

Furthermore, no large differences in MAE were observed when using a different number of latent channels for the $2\times2$ and $8\times8$ models (Figure B.1 in Appendix B). However, the models with 32 latent channels showed the highest degree of overlap between MOOD\textsubscript{val} and MOOD\textsubscript{train}, and the lowest degree of overlap between MOOD\textsubscript{val} and MOOD\textsubscript{art\_circ}. Therefore, 32 latent channels were used for further experiments.

\subsection{Optimal versus final epoch}
\begin{figure}
    \centering
    \includegraphics[width=\linewidth]{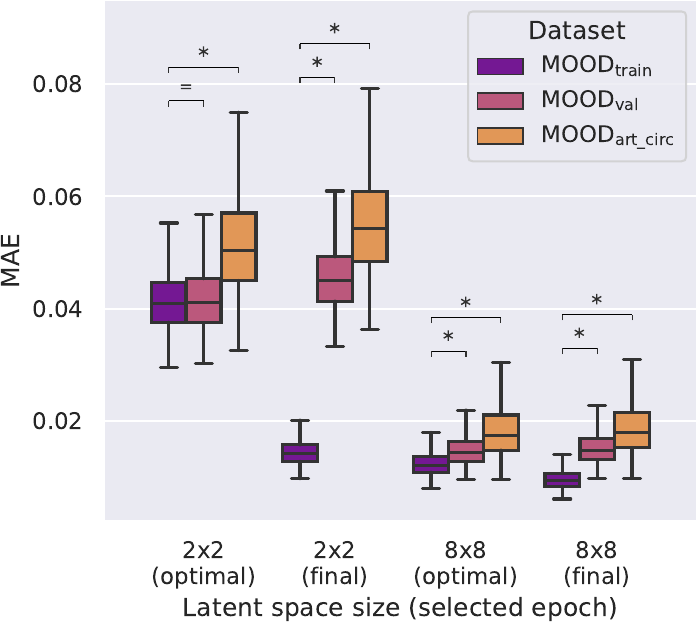}
    \caption{MAE of the reconstructions obtained for different selected epochs and models. Statistical significance (\(\alpha=0.01\)) was assessed between MOOD\textsubscript{train} and MOOD\textsubscript{val}, as well as between MOOD\textsubscript{train} and MOOD\textsubscript{art\_circ}. Significant differences are indicated by an asterisk and non-significant differences are indicated by an equal sign. Note that outliers are not shown in the boxplots for visibility purposes.}
    \label{fig:boxplots_best_last}
\end{figure}

\begin{figure}
    \centering
    \includegraphics[width=\linewidth]{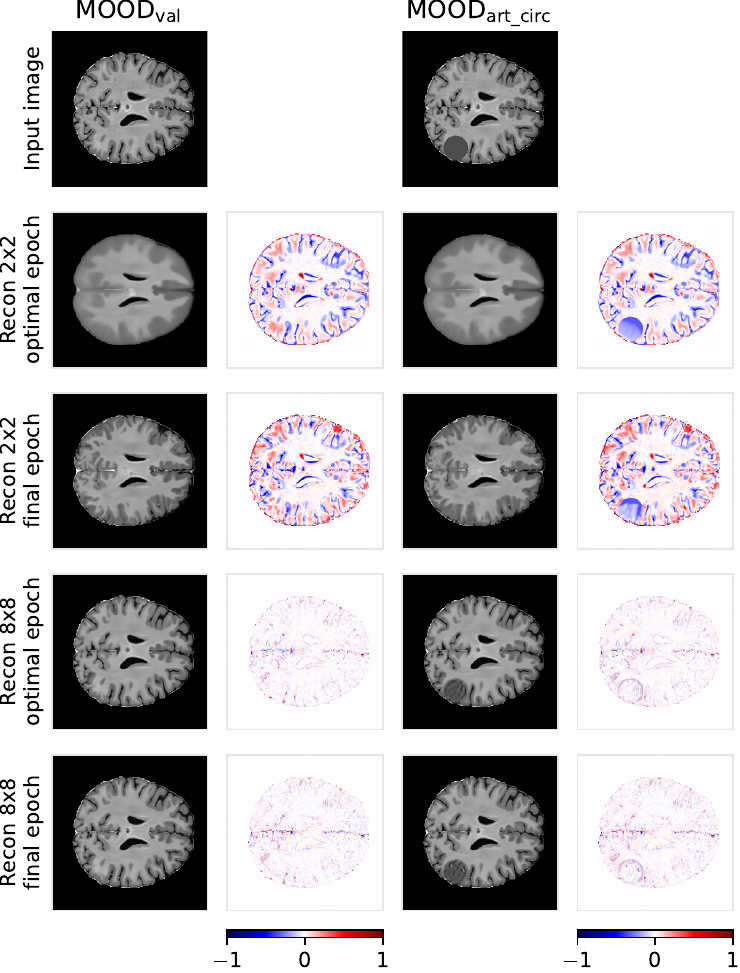}
    \caption{Visual results based on different selected epochs and models. The first row shows the same example as in \autoref{fig:recons_latent_sizes} from MOOD\textsubscript{val} and its counterpart MOOD\textsubscript{art\_circ}. Second and third rows show the reconstructions and errors ($x-\hat{x}$) of the $2\times2$ model based on the optimal and final epoch, respectively. Fourth and fifth rows show the reconstructions and errors ($x-\hat{x}$) of the $8\times8$ model based on the best and final epoch, respectively.}
    \label{fig:recons_best_last}
\end{figure}

A general expectation is that when a deep learning model is trained for an extended period, it will overfit to the training data. This was indeed observed by examining the reconstruction errors of our models (\autoref{fig:boxplots_best_last}). Specifically for the $2\times2$ model, the model shows strong overfitting to the MOOD\textsubscript{train} set and does not generalize well to the MOOD\textsubscript{val} set when the last epoch was selected compared to when the epoch with the lowest validation loss was selected. However, when analyzing two examples in \autoref{fig:recons_best_last}, the reconstruction appears to be of higher quality when the final epoch is selected compared to the optimal epoch. It seems that the anatomy of the brain has changed as a result of the reconstruction based on the final epoch, which also resulted in a large absolute error similar to that of the optimal epoch. In the case of MOOD\textsubscript{art\_circ}, the anomaly is healed in addition to the changed brain anatomy. Thus, at first glance, one would assume that the anomalous region would be detected, but this region is not significantly out of distribution compared to the changed brain structures and therefore not clearly identified when using the absolute error as reconstruction metric.

\subsection{Choice of metric}
\begin{figure*}
    \centering
    \includegraphics[width=\linewidth]{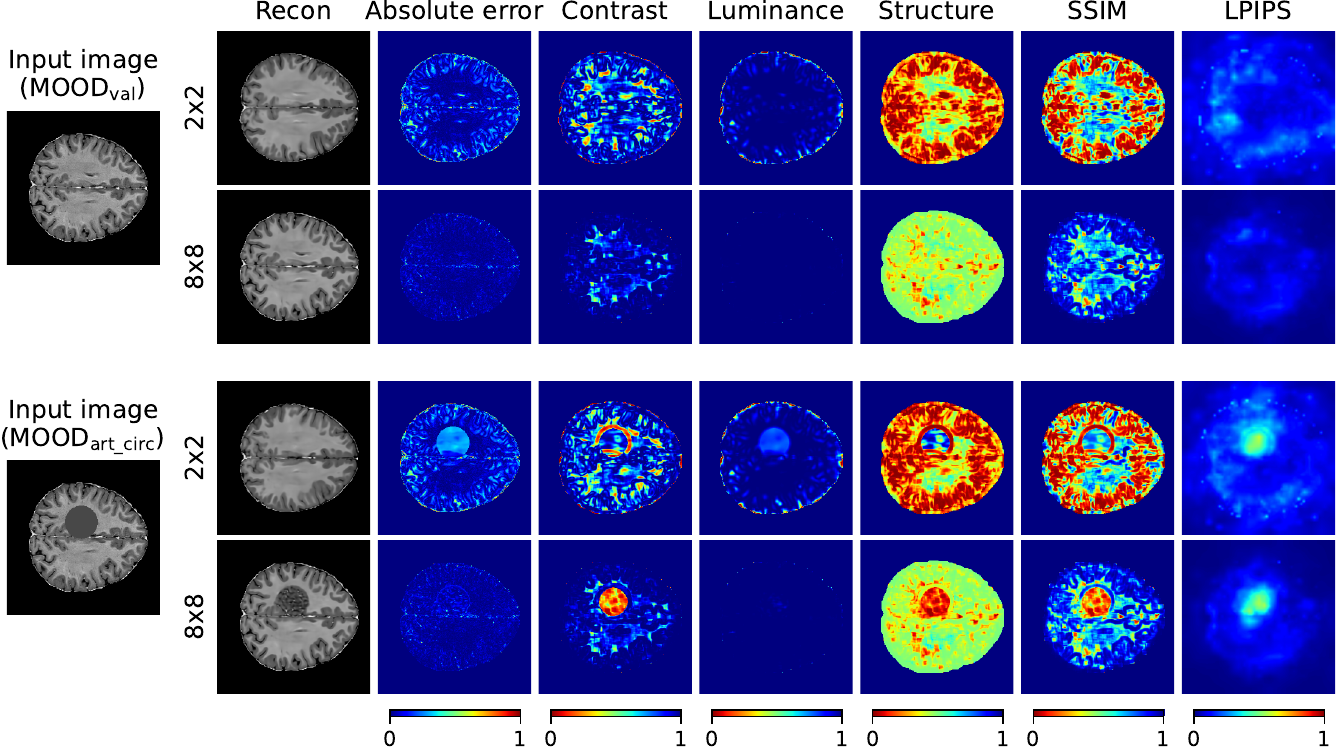}
    \caption{Randomly selected image from MOOD\textsubscript{val} and its OOD counterpart from MOOD\textsubscript{art\_circ} (first column) with the corresponding reconstructions (second column) and the corresponding reconstruction errors (third to last columns). The first two rows show the results of the MOOD\textsubscript{val} image for the $2\times2$ and $8\times8$ models, respectively, and the last two rows show the results of the MOOD\textsubscript{art\_circ} image for the $2\times2$ and $8\times8$ models, respectively. }
    \label{fig:metric_maps}
\end{figure*}

\begin{figure}
    \centering
    \includegraphics[width=\linewidth]{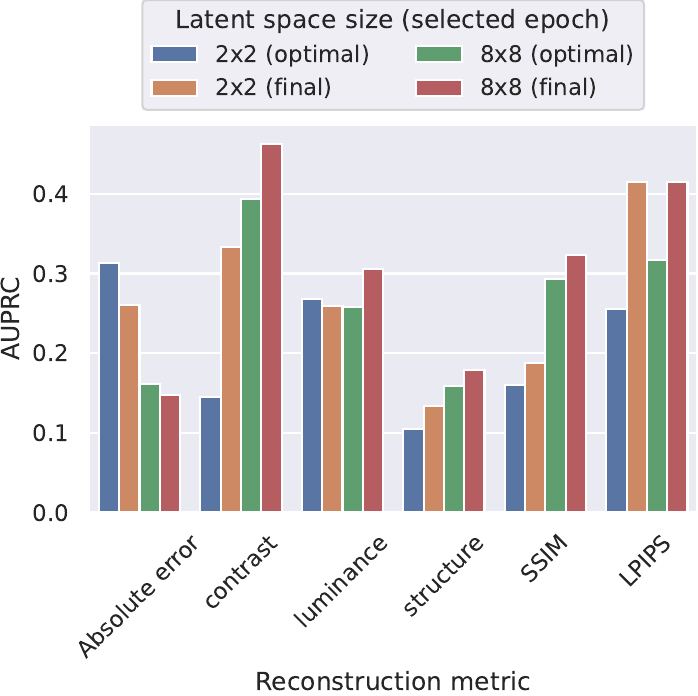}
    \caption{Average precision of MOOD\textsubscript{art\_circ} pixel-level predictions based on six different metrics: MAE, SSIM, contrast, luminance, structure, and LPIPS.}
    \label{fig:bars_AP}
\end{figure}

\autoref{fig:metric_maps} highlights the impact of metric selection on detecting anomalies in reconstructed images, demonstrating that each metric reveals different reconstruction characteristics. The absolute error for the $2\times2$ model show small deviations across the cortical folds, including both gray and white matter, while the central brain regions show limited errors. In contrast, the absolute error for the $8\times8$ model shows minimal deviations overall, with neither model effectively capturing the abnormality. The contrast component effectively detects the anomaly in both models, although the $2\times2$ model includes more false positive structures throughout the brain. Luminance errors remain generally low for both models, with the $2\times2$ model showing minor errors at the outer brain boundary. The structure component reveals significant reconstruction errors in the cortical folds for the $2\times2$ model. Based on the structure, the reconstruction quality of the $8\times8$ model is relatively low but consistent, which allows better detection of the anomaly. 
Looking at the SSIM, we observe that the low performance of the structure component overshadows the contrast and luminance components. SSIM again highlights large errors within the cortical folds for the $2\times2$ model, with reduced errors for the $8\times8$ model, where the anomaly is more pronounced. Lastly, LPIPS presents low errors with a smooth appearance, with only slight emphasis on the outer region of the brain in the $2\times2$ model and no effective capture of the brain structure or anomaly. In terms of quantitative analysis (\autoref{fig:bars_AP}), the contrast metric demonstrated the highest AUPRC of 0.46 for the final epoch of the $8\times8$ model. Furthermore, the LPIPS metric demonstrated a high AUPRC value of 0.42 for the final epochs of both the $2\times2$ model and the $8\times8$ model. These findings suggest that metrics sensitive to perceptual errors, such as LPIPS, and those focusing on contrast variations, such as the contrast metric, may provide more effective anomaly localization.

\subsection{Ensemble of models and metrics}
\begin{figure*}
    \centering
    \includegraphics[width=\linewidth]{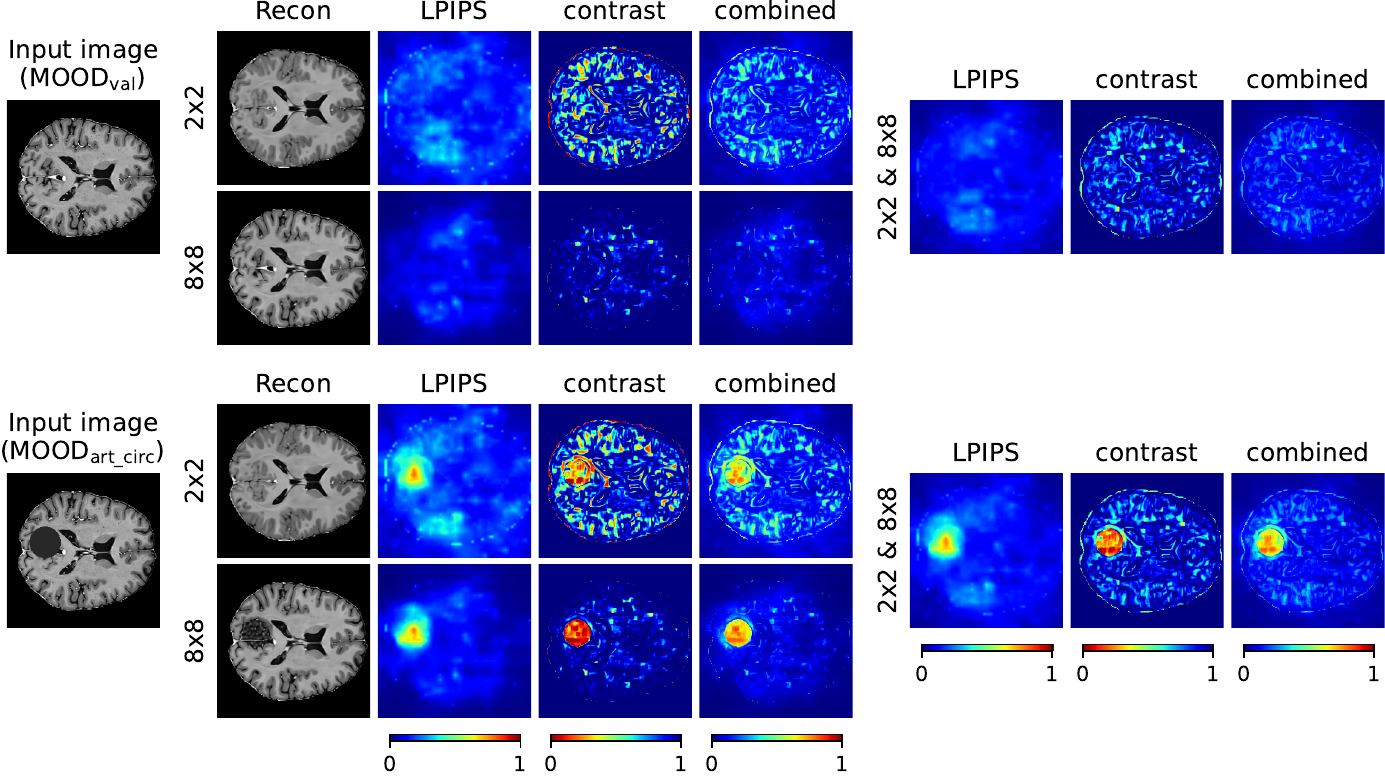}
    \caption{Randomly selected input image and its OOD counterpart, accompanied by their reconstructed images and prediction maps. The different columns represent the 1) input images, 2) reconstructions from a single model, 3) LPIPS from a single model, 4) contrast from a single model, 5) prediction maps combining LPIPS and contrast from a single model, 6) LPIPS from both models averaged, 7) contrast from both models averaged, and 8) prediction maps combining LPIPS and contrast from both models.}
    \label{fig:visuals_ensembled}
\end{figure*}
\autoref{fig:visuals_ensembled} shows a randomly selected input image alongside its OOD counterpart, along with the reconstructions derived from the $2\times2$ and $8\times8$ models. The figure includes separate prediction maps for LPIPS and contrast, as well as combined predictions. Notably, the combination of LPIPS and contrast reduces false positive structures in the brain without substantially compromising anomaly detection. The LPIPS metric primarily identifies the anomaly from the $2\times2$ model, while the $8\times8$ model provides a more subtle indication. In contrast, the $8\times8$ model clearly highlights the anomaly with fewer false positive structures compared to the $2\times2$ contrast prediction, which is marked by additional false detections in other brain regions. This observation suggests that combining models and metrics effectively leverages the strengths of both models and metrics, optimizing anomaly detection while reducing false positives.

\begin{figure*}
    \centering
    \includegraphics[width=\linewidth]{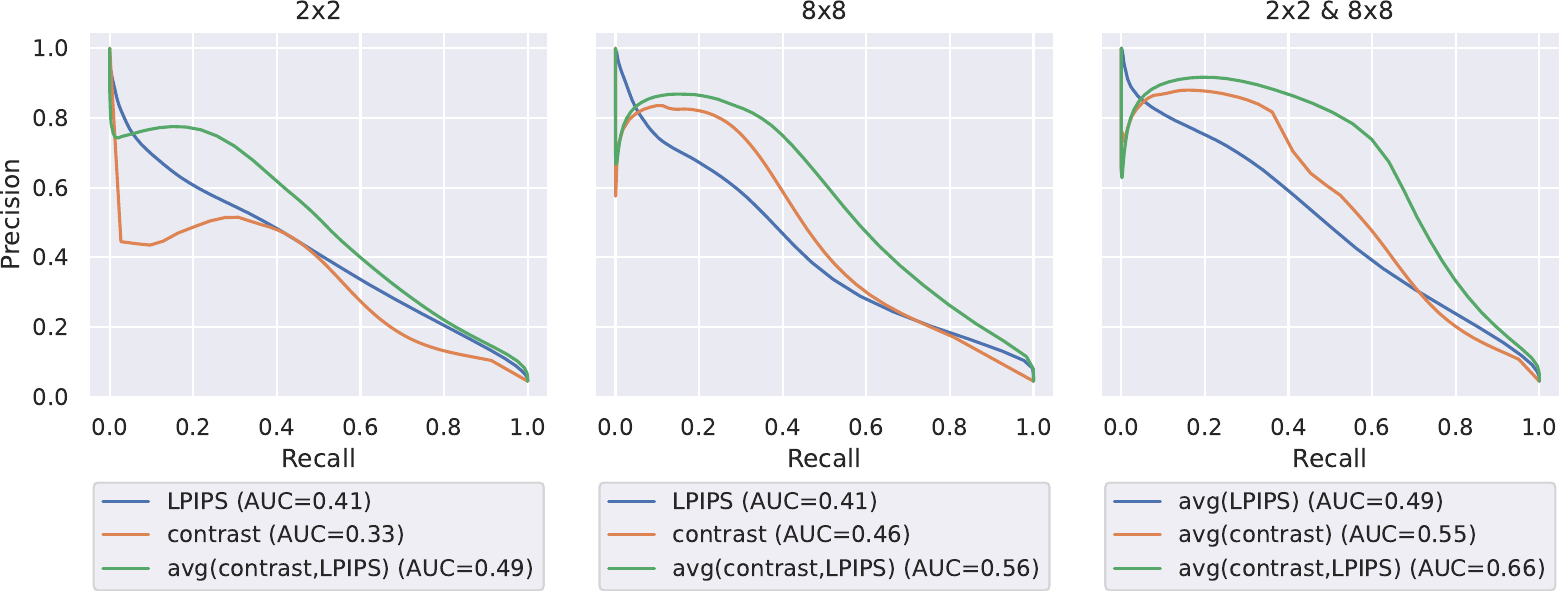}
    \caption{Precision--Recall curves for MOOD\textsubscript{art\_circ} based on LPIPS, contrast, and combined prediction maps for from left to right: $2\times2$ model, $8\times8$ model, and both models combined.}
    \label{fig:PR_curves_ensembling}
\end{figure*}

The prediction map that combines both models and both metrics results in a visually coherent and perceptually pleasing representation. This visual observation is further supported by the AUPRC for the individual and combined reconstruction maps (\autoref{fig:PR_curves_ensembling}). It can be observed that the combined prediction map resulted in a higher AUPRC than when each model was considered separately. Specifically, for the $2\times2$ model, the combination of LPIPS (AUPRC = 0.71) and contrast (AUPRC = 0.33) resulted in an AUPRC of 0.49. Similarly, for the $8\times8$ model, the combination of LPIPS (AUPRC = 0.41) and contrast (AUPRC = 0.46) resulted in an AUPRC of 0.56. Furthermore, when combining both models per metric, there was an increase in AUPRC when using the LPIPS (AUPRC = 0.49) and contrast (AUPRC = 0.55). Moreover, an even more pronounced improvement in performance was observed when the predictions of both models and both metrics were combined, resulting in an AUPRC of 0.66.

\subsection{Extended synthetic artifact dataset}
\begin{figure}
    \centering
    \includegraphics[width=\linewidth]{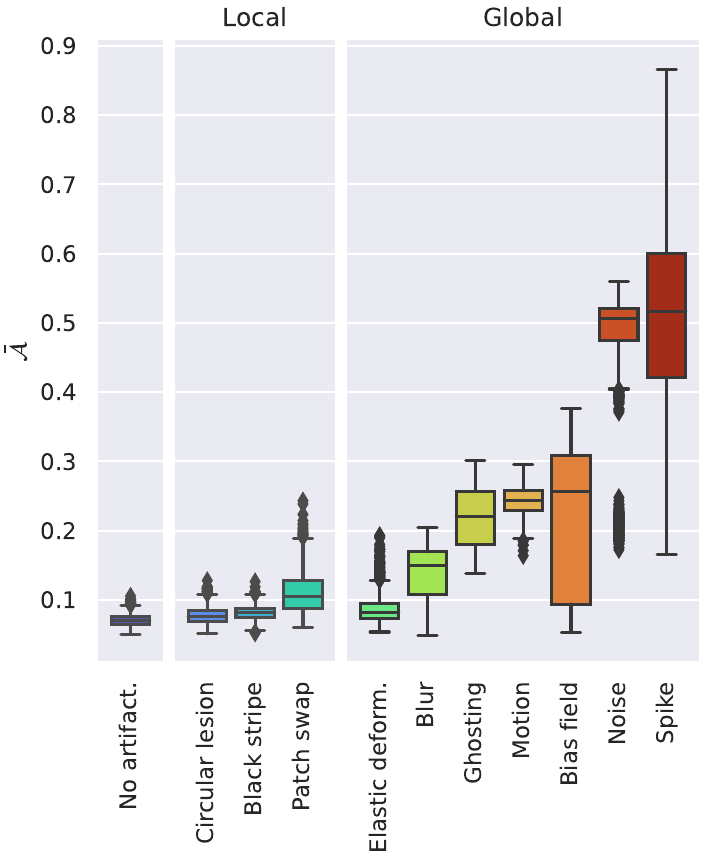}
    \caption{The anomaly score \(\bar{\mathcal{A}}\) for each slice for the MOOD\textsubscript{val} dataset (first boxplot) and for each artifact type in MOOD\textsubscript{art\_ext} (second to last boxplots). All boxplots show a significant higher \(\bar{\mathcal{A}}\) compared to MOOD\textsubscript{val}.}
    \label{fig:boxplots_transformed}
\end{figure}

All artifact types resulted in a significantly higher \(\bar{\mathcal{A}}\) compared to the \(\bar{\mathcal{A}}\) of MOOD\textsubscript{val} (\(p \sim 0.00\)), indicating that all artifact types resulted in a measurable deviation in OOD detection (\autoref{fig:boxplots_transformed}). Although the circular lesion and black stripe artifacts had only slightly elevated mean scores of 0.078 and 0.082, respectively, compared to the ID validation mean of 0.070, paired analysis showed that they had consistently higher scores across all paired samples. As mentioned above, the global performance \(\mathcal{\bar{A}}\) for the local artifacts is highly dependent on the size of the anomaly, which is not the case when evaluating at pixel-level. AUPRC values of 0.50, 0.18, and 0.55 were found for the subsets of circular lesions, black stripes, and patch swaps, respectively. Contrary to the observations presented in \autoref{fig:boxplots_transformed}, the AUPRC indicates that detecting black stripes is more difficult than detecting the circular lesions and patch swaps.

High values for \(\mathcal{\bar{A}}\) were found for ghosting, motion, noise, and spikes (\autoref{fig:boxplots_transformed}). However, for these artifacts, the background pixel intensities were altered due to the design of the artifact generation method. This led to large reconstruction errors in the background regions, resulting in high values for \(\mathcal{\bar{A}}\) (Figure B.2 in Appendix B). In addition, the bias field artifact shows a large spread in \(\bar{\mathcal{A}}\) and was also generally well captured by the model. Nevertheless, a large spread in the values of \(\mathcal{\bar{A}}\) is not unexpected since the datasets were generated by systematically increasing the values of the artifact parameters.

\begin{figure}
    \centering
    \includegraphics[width=\linewidth]{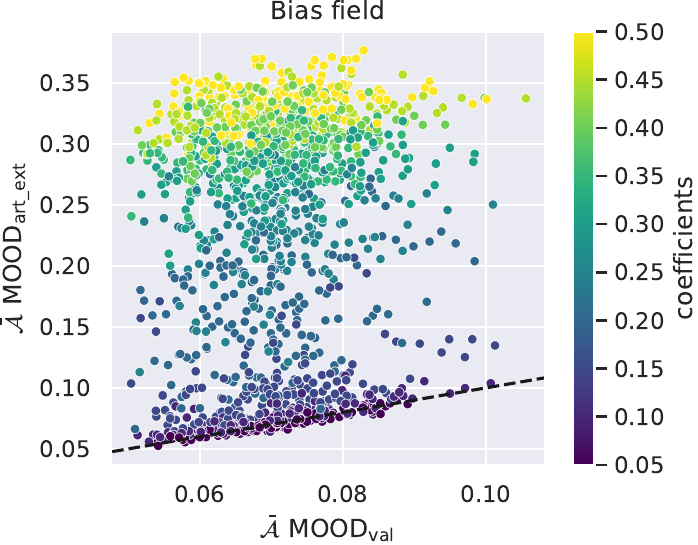}
    \caption{The score \(\bar{\mathcal{A}}\) for all of bias field slices in MOOD\textsubscript{art\_ext} versus \(\bar{\mathcal{A}}\) of the same slice in the MOOD\textsubscript{val} dataset. The colors indicate the different coefficients used to the create the bias field cases. The black dashed line indicates that \(\bar{\mathcal{A}}\) MOOD\textsubscript{art\_ext} is equal to \(\bar{\mathcal{A}}\) MOOD\textsubscript{val}.}
    \label{fig:corr_biasfield}
\end{figure}

\begin{figure}
    \centering
    \includegraphics[width=\linewidth]{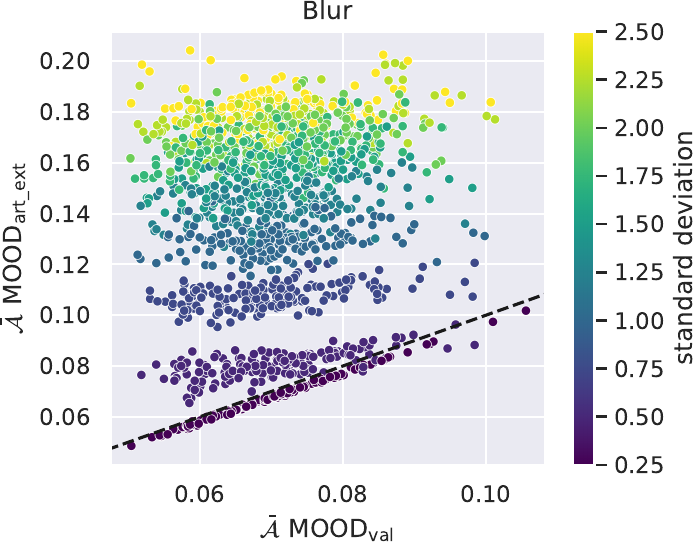}
    \caption{The score \(\bar{\mathcal{A}}\) for all of blurred slices in MOOD\textsubscript{art\_ext} versus \(\bar{\mathcal{A}}\) of the same slice in the MOOD\textsubscript{val} dataset. The colors indicate the different standard deviations used to the create the blurred cases. The black dashed line indicates that \(\bar{\mathcal{A}}\) MOOD\textsubscript{art\_ext} is equal to \(\bar{\mathcal{A}}\) MOOD\textsubscript{val}.}
    \label{fig:corr_blur}
\end{figure}

\begin{figure*}
    \centering
    \includegraphics[width=\linewidth]{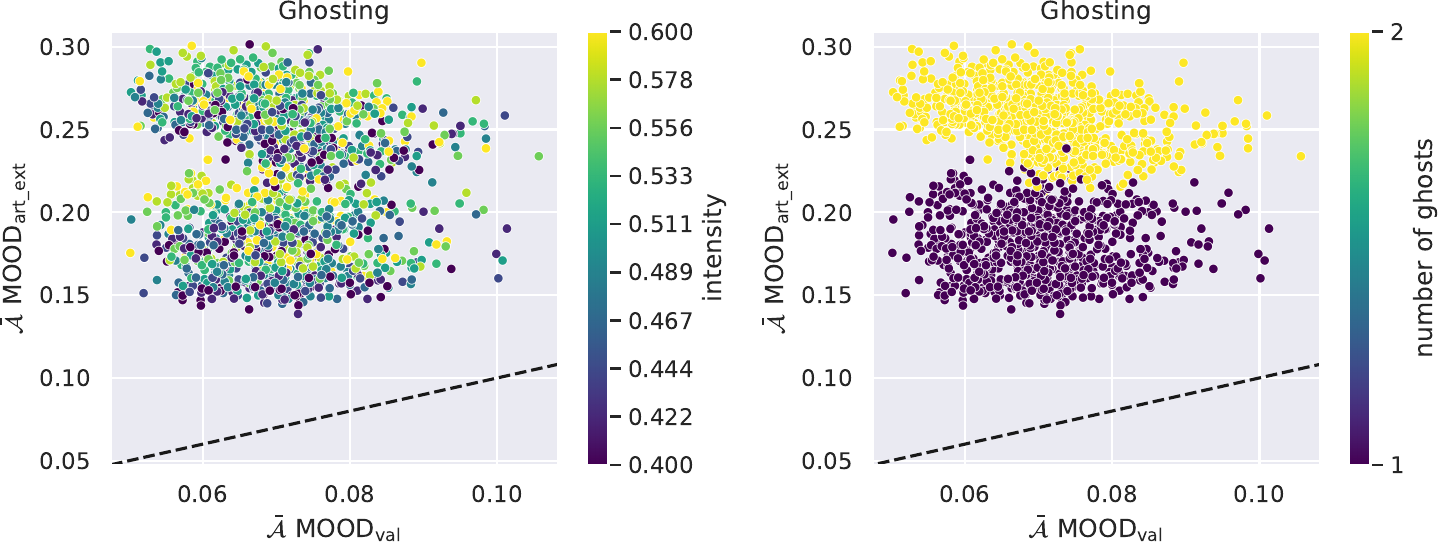}
    \caption{The score \(\bar{\mathcal{A}}\) for all ghosting slices in MOOD\textsubscript{art\_ext} versus \(\bar{\mathcal{A}}\) of the same slice in the MOOD\textsubscript{val} dataset. The colors in the left and right figures indicate the different intensities and number of ghosts, respectively, used to the create the cases with ghosting. The black dashed line indicates that \(\bar{\mathcal{A}}\) MOOD\textsubscript{art\_ext} is equal to \(\bar{\mathcal{A}}\) MOOD\textsubscript{val}.}
    \label{fig:corr_ghost}
\end{figure*}

The expectation that a higher severity of the artifact would lead to a higher value of \(\mathcal{\bar{A}}\) was confirmed for most artifact types. This is particularly evident for the blur (\autoref{fig:corr_blur}) and bias field (\autoref{fig:corr_biasfield}). The standard deviation used for blurring and the coefficients used for the bias field show near-perfect correlations with \(\bar{\mathcal{A}}\) of MOOD\textsubscript{art\_ext}, with Spearman rank correlation coefficients of \(\rho = 0.962\) and \(\rho = 0.956\), respectively (\autoref{tab:spearman}). In contrast, other artifact types showed weaker correlations. For example, for the ghosting artifact there was a strong correlation (\(\rho = 0.864\)) between the number of ghosts and the score \(\bar{\mathcal{A}}\) of MOOD\textsubscript{art\_ext}, while the intensity showed significantly less impact on the model's detection capability, with a correlation of only \(\rho = 0.184\). This trend is also reflected in the scatter plots presented in \autoref{fig:corr_ghost}.

A notable observation is the correlation of $\rho = -0.123$ between the intensity of circular lesions and $\mathcal{\bar{A}}$, which stands out as the only negative correlation across all artifact types and the one with the smallest magnitude. A large positive \(\rho\) would suggest that brighter circular lesions are better detected, while a large negative \(\rho\) would imply that darker lesions are easier to detect. However, the observed weak negative correlation suggests the absence of a strong linear correlation. This could indicate either that there is no relationship or that detection is influenced by non-linear factors, such as the contrast of the lesion relative to the surrounding tissue. One could speculate that circular lesions with extreme intensity values (very dark or very bright) may be better detected because of the higher contrast. Figure B.3 in Appendix B provides further insight into this relationship, showing no strong overall correlation between intensity and \(\mathcal{\bar{A}}\). Nevertheless, very dark lesions with low intensities appear to stand out slightly, demonstrating marginally better detection compared to other intensity values. 

Another notable observation, based on \autoref{fig:boxplots_transformed} and \autoref{tab:spearman}, is that while motion artifact is well detected as anomaly, the correlations between the detected motion and the motion parameters are relatively weak, with values of \(\rho = 0.241\) and \(\rho = 0.091\) for rotation and translation, respectively. This indicates that the amount of rotation and translation does not substantially affect the detection process.

Scatter plots for other artifact types not shown in the main paper can be found in Appendix B, providing a comprehensive overview of these correlations across all artifact parameters.

\begin{table}
\caption{Spearman rank correlation coefficient \(\rho\) between the anomaly score \(\bar{\mathcal{A}}\) of MOOD\textsubscript{art\_ext} and the artifact parameter.}
{\setlength{\arrayrulewidth}{0.1mm} 
 \renewcommand{\arraystretch}{1.2}  
\begin{tabular}{lll}
\toprule
\textbf{Local artifact} & \textbf{Parameter} & \textbf{$\rho$} \\

\midrule
\multirow{2}{*}{Circular lesion} & radius & 0.518 \\
 & intensity & -0.123 \\

\hline
Black stripe & thickness & 0.124 \\
 
\hline
{Patch swap} & patch size & 0.599 \\
 
\bottomrule
\toprule
\textbf{Global artifact} & \textbf{Parameter} & \textbf{$\rho$} \\

\midrule
\multirow{2}{*}{Elastic deformation} & maximum displacement & 0.581 \\
 & number of control points & 0.356 \\
 
 \hline
Blur & standard deviation & 0.962 \\

\hline
\multirow{2}{*}{Ghosting} & intensity & 0.184 \\
 & number of ghosts & 0.864 \\

\hline
\multirow{2}{*}{Motion} & rotation & 0.241 \\
 & translation & 0.091 \\

 \hline
Bias field & coefficients & 0.956 \\

\hline
{Noise} & standard deviation & 0.745 \\

\hline
\multirow{2}{*}{Spike} & intensity & 0.808 \\
 & number of spikes & 0.257 \\
\bottomrule
\end{tabular}
}
\label{tab:spearman}
\end{table}

%% file: 07_Discussion.tex
\section{Discussion}
\label{OODsec:discussion}

The detection of OOD and anomalies in medical images remains a challenging and unsolved area of research. While numerous pipelines have been developed, many are overly tailored to specific datasets, limiting their generalizability. In addition, we observe a trend to increase model complexity, while the potential of less complex models remains underexplored. Another important research direction is the exploration of deep learning strategies and reconstruction metrics explicitly tailored for OOD detection. In this study, we focused on OOD detection in brain MRI using an reconstruction-based AE, emphasizing the importance of deep learning design choices and metric selection.

This study shows that redefining the OOD detection pipeline, specifically the choice of reconstruction metric, in a standard AE can improve performance compared to using a standard OOD detection pipeline with an absolute error reconstruction metric. Our model with the smallest latent space size achieved the best separation between ID and OOD data while maintaining generalizability, whereas models with larger latent spaces reconstructed both ID and OOD data well. These results are consistent with previous research that emphasizes the importance of balancing latent space size and reconstruction capability in OOD detection \citep{baur2021autoencoders, cai2024rethinking}. In addition, our work highlights the critical yet underexplored role of selecting the training epoch and the reconstruction metric in OOD detection. We found that the optimal training epoch for the reconstruction model does not always coincide with the best detection performance, and that metrics sensitive to perceptual errors and those focusing on contrast variations may provide more effective anomaly localization. By selecting a well-converged model and using appropriate metrics, we improved OOD detection performance. Specifically, the AUPRC for the $2\times2$ and $8\times8$ models based on the absolute error was 0.26 and 0.15, respectively, while combining both models and using the LPIPS and contrast metrics resulted in an AUPRC of 0.66. Direct comparisons with existing literTchallenging due to the lack of benchmark data. However, the cl \citep{zimmerer2022mood}osest comparison can be made with the MOOD 2020 challeng circular artifact dataset. In this challenge, the median ranked team achieved a pixel-level AUPRC of approximately 0.77 on the toy test set. However, this value reflects the influence of various factors, such as the larger anomaly sizes (up to twice the radius of our maximum radius) and the possibility that teams optimized their models to detect such homogeneous toy-like anomalies, while our model was not specifically tuned for this purpose. These differences prohibit a proper comparison with our AUPRC of 0.66.

Among the evaluated metrics, contrast and LPIPS consistently outperformed others in detecting circular artifacts. The strong performance of LPIPS is consistent with previous studies highlighting the usefulness of perceptual metrics in anomaly detection \citep{bercea2023generalizing}. Other studies have also explored the potential of SSIM ensembles derived from multiple resolutions, features, or implementations \citep{tian2023unsupervised,meissen2022unsupervised,behrendt2024diffusion}. However, our study investigates the individual components of SSIM for OOD detection, providing a novel perspective on this widely used metric.

We developed our proposed pipeline using a dataset with homogeneous circular artifacts and applied it directly to an extended dataset with more artifact types without any fine-tuning. The results showed variability in performance across artifact types, depending on the nature of the artifact. Local artifacts were significantly more difficult to detect, yielding AUPRC scores ranging from 0.18 to 0.55, compared to an AUPRC of 0.66 on the circular artifact dataset. Although most of the global artifacts were clearly distinguishable from the ID validation set, the severity of some artifacts significantly affected how OOD a sample appeared. This raises a question for OOD detection models in medical imaging: at what point should an image be flagged as OOD, or when might it be better categorized as rare but still ID?

Our study has several limitations that may affect the interpretation and generalizability of the results. First, the synthetic nature of our OOD datasets may reduce the applicability to real-world clinical scenarios, where medical anomalies exhibit different characteristics. Additionally, the extended artifact dataset may have caused misleading results due to changes in background intensities in some of the global artifacts caused by data re-normalization and could have been corrected for. These background intensity changes made it challenging to interpret the correlation between artifact parameters and detection performance. Another limitation is our primary focus on pixel-level anomaly detection, leaving binary sample-level assessments unexplored. Given that our extended artifact dataset consisted primarily of global artifacts, exploring sample-level detection could have provided valuable insights. However, we chose not to pursue this because it often requires extensive tuning of post-processing and thresholding pipelines to a specific OOD set. 

Our findings highlight several promising directions for future research. First, a more detailed analysis of model convergence and overfitting for OOD detection is recommended. Such an analysis could, for example, include nearest-neighbor comparisons between test samples and the training set to better understand how overfitting affects the representation of healthy tissue in OOD test samples. Second, validation of the proposed approach on clinical OOD datasets is essential. Such datasets would capture a variety of medical anomalies and/or scanning artifacts, allowing for a more comprehensive evaluation of the model's performance. The nature of the anomalies (local or global) and the intended application of OOD detection (e.g., anomaly localization or flagging of OOD samples for a downstream task) will determine the need to extend the sample-level prediction pipeline to enhance its utility. A final direction for future work is to investigate how the proposed metric ensemble generalizes to other reconstruction-based approaches, such as (latent) diffusion models for medical OOD detection \citep{wyatt2022anoddpm,bercea2023mask,wolleb2024binary}. These models have shown strong performance and could benefit from the insights gained through our metric ensemble approach.

In conclusion, our study demonstrates that a straightforward reconstruction-based AE, when combined with perceptually sensitive metrics such as contrast and LPIPS, can effectively improve OOD detection in brain MRI. We highlight the importance of both latent space size and metric selection for accurate detection.  Additionally, our findings demonstrate that different deep learning design choices, such as model depth, number of training epochs, and metric selection, must be carefully evaluated for the task of OOD detection, as standard approaches often originate from different objectives and may fail to capture various types of anomalies.

%% file: 09_Appendix.tex
\noindent This document is a supplementary file to Huijben et al. ``Enhancing Reconstruction-Based Out-of-Distribution Detection in Brain MRI with Model and Metric Ensembles''. In this supplementary document, we provide a detailed description of the creation of our synthetic test sets in \autoref{OODsec:appendix_dataset} and more detailed results and analyses of the extended artifact dataset in \autoref{OODsec:appendix_results}.

\section{dataset details}
\label{OODsec:appendix_dataset}

The code for creating MOOD\textsubscript{art\_circ} and MOOD\textsubscript{art\_ext} is available at \url{https://github.com/evihuijben/ood_detection_mri}.

\subsection{Validation set with circular artifacts}
\label{OODsec:appendix_circ_dataset}
To create the MOOD\textsubscript{art\_circ}, the 2D slices were loaded and the artifact was also added in 2D. The circular artifacts were created by selecting a random center point within the specified body mask for each slice. This body mask includes all non-zero pixels within the input slice, excluding pixels located at a distance of less than radius $r$ from the border. A black image with a white circle of radius r and center coordinate was then created using the \texttt{raster-geometry.circle} module\footnote{\url{https://github.com/norok2/raster_geometry}}. To obtain the final brain slice with the circular artifact, the following equation was used: \(x \odot (1-C) + i \cdot C\), where $x$ is the input slice, $C \in \mathbb{R}^{256\times256}$ is the circle image, and $i$ is the intensity. The ground truth masks used for the local evaluation were equal to the circle image $C$.

\subsection{Extended synthetic artifact dataset}
We introduced three types of local artifacts: circular lesion, black stripe, and patch swap. In addition, we introduced seven types of global artifacts: blurring, noise, elastic deformation, patient motion, MRI bias field, MRI ghosting artifact, and MRI spikes (also known as Herringbone artifact). Note that before applying the transformation, the 3D volumes were normalized between zero and one. The black stripe and circular lesion artifacts were custom artifacts, and all other artifacts were based on the \texttt{torchio.transforms} library \citep{perez2021torchio}, as described in the following sections.

\subsubsection{Local artifacts} 
To create the local artifacts, the 2D slices were loaded and the transformation was also applied in 2D. The ground truth masks utilized for the local evaluation included every pixel that was altered by the transformation.

\textbf{\emph{Circular lesions}} were created similarly to MOOD\textsubscript{art\_circ} as described in \autoref{OODsec:appendix_circ_dataset}. However, as described in Table 1, the values of the range of radii $r$ of the circular lesions in MOOD\textsubscript{art\_ext} were smaller than the values of the range of radii $r$ for MOOD\textsubscript{art\_circ}. In addition, the smoothing parameter of the \texttt{raster-geometry.circle} module was set to true for the circular lesions, resulting in smoother edges for the circles in MOOD\textsubscript{art\_ext} compared to those in MOOD\textsubscript{art\_circ}.

\textbf{\emph{Black stripe}} was generated by selecting a random column/row of pixels between the first and last column/row of pixels that included non-zero intensities. A black stripe with a randomly sampled orientation (that is, a column or a row) and a predefined thickness $N$, was created for each input slice.

\textbf{\emph{Patch swap}} was developed using an adapted version of \texttt{torchio.transforms.RandomSwap}, where a patch must contain at least 10\% of foreground pixels (i.e., non-zero pixels) in order for it to be eligible for swapping. A single iteration of the swapping process was performed and due to the 2D implementation, the patch size was defined as $w\times h\times1$, where $w$ represents the width, $h$ represents the height, and $w=h$.

\subsubsection{Global artifacts} 
To create the global artifacts per 2D slice, the 3D volumes were loaded, the transformation was applied in 3D, the transformed 3D volume was normalized between zero and one, and only the designated 2D slice was saved.

\textbf{\emph{Blurring}} was implemented using \texttt{torchio.transforms.RandomBlur}, and to ensure that the standard deviation was set to the desired value of $\sigma$, we passed a tuple of (\(\sigma, \sigma\)) to the function.

\textbf{\emph{Noise}} was implemented using \texttt{torchio.transforms.RandomNoise} with a mean of 0. To ensure that the standard deviation was set to the desired value of $\sigma$, we passed a tuple of (\(\sigma, \sigma\)) to the function.

\textbf{\emph{Elastic deformation}} was implemented using \texttt{torchio.transforms.RandomElasticDeformation} with the desired number of control points along each dimension $n$ and the desired maximum displacement $d$.

\textbf{\emph{Patient motion}} was developed using an adapted version of \texttt{torchio.transforms.RandomMotion}, in which we replaced the random selection of a rotation and translation axis with a fixed axis. We fixed the rotation axis to the inferior-superior axis and randomly selected the translation axis to be either the left-right axis or the anterior-posterior axis. A single motion transformation that included both rotation and translation was applied and a positive or negative rotation and translation of the specified degrees and translation, respectively, was selected.

\textbf{\emph{Bias field}} was implemented using \texttt{torchio.transforms.RandomBiasField}, and to ensure that the maximum magnitude $n$ of polynomial coefficients was set to the desired value, we passed a tuple of (\(n, n\)) to the function.

\textbf{\emph{Ghosting}} was implemented using \texttt{torchio.transforms.RandomGhosting}, with either one or two ghosts and in the left-right or superior-inferior direction. Furthermore, to ensure that the strength (i.e., intensity) of the ghost(s) was set to the desired value of $i$, we passed a tuple of (\(i,i\))  to the function.

\textbf{\emph{Spike}} was developed using an adapted version of \texttt{torchio.transforms.RandomSpike}, in which we replaced the random sampling for the intensity value with a fixed value. Specifically, the ratio between the spike intensity and the maximum of the spectrum, was fixed to the positive or negative value of the passed intensity $i$.

\section{Supplementary results}
\label{OODsec:appendix_results}
\subsection{Reconstruction capability}
\label{OODsec:appendix_results_recon}

\begin{figure*}[h]
    \centering
    \includegraphics[width=\linewidth]{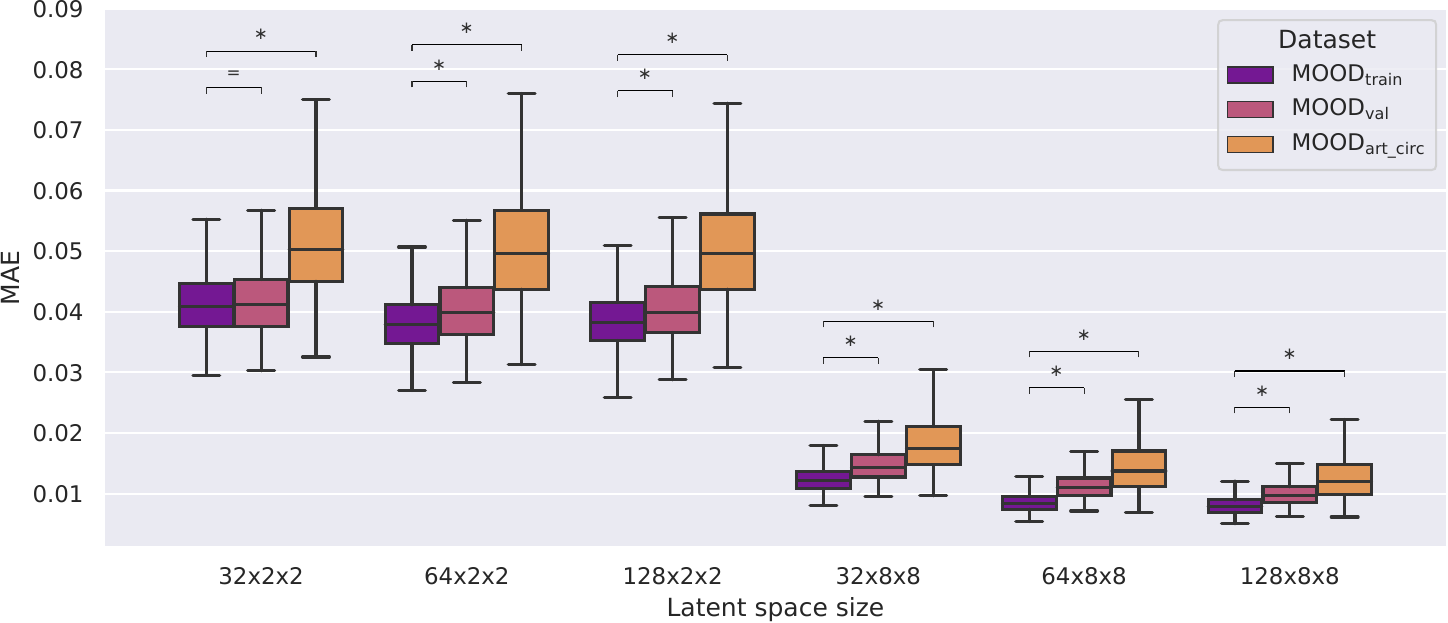}
    \caption{MAE of the AE with a spatial latent space size of $2\times2$ and $8\times8$ and varying number of latent channels. Statistical significance (\(\alpha=0.01\)) was assessed between MOOD\textsubscript{train} and MOOD\textsubscript{val}, as well as between MOOD\textsubscript{train} and MOOD\textsubscript{art\_circ}. Significant differences are indicated by an asterisk and non-significant differences are indicated by an equal sign. Note that outliers are not shown in the boxplots for visibility purposes.}
    \label{fig:box_latent_channels}
\end{figure*}

This subsection presents additional results for the reconstruction capability of the $2\times2$ and $8\times8$ models with latent channel configurations of 32, 64, and 128. As shown in \autoref{fig:box_latent_channels}, no significant differences in MAE were observed between the varying latent channels for these models. However, the models with 32 latent channels showed the highest degree of overlap between MOOD\textsubscript{val} and MOOD\textsubscript{train}, and the lowest degree of overlap between MOOD\textsubscript{val} and MOOD\textsubscript{art\_circ}. 

\newpage
\subsection{Extended synthetic artifact dataset}
\label{OODsec:appendix_results_transformed}

\begin{figure*}[h]
    \centering
    \includegraphics[width=\linewidth]{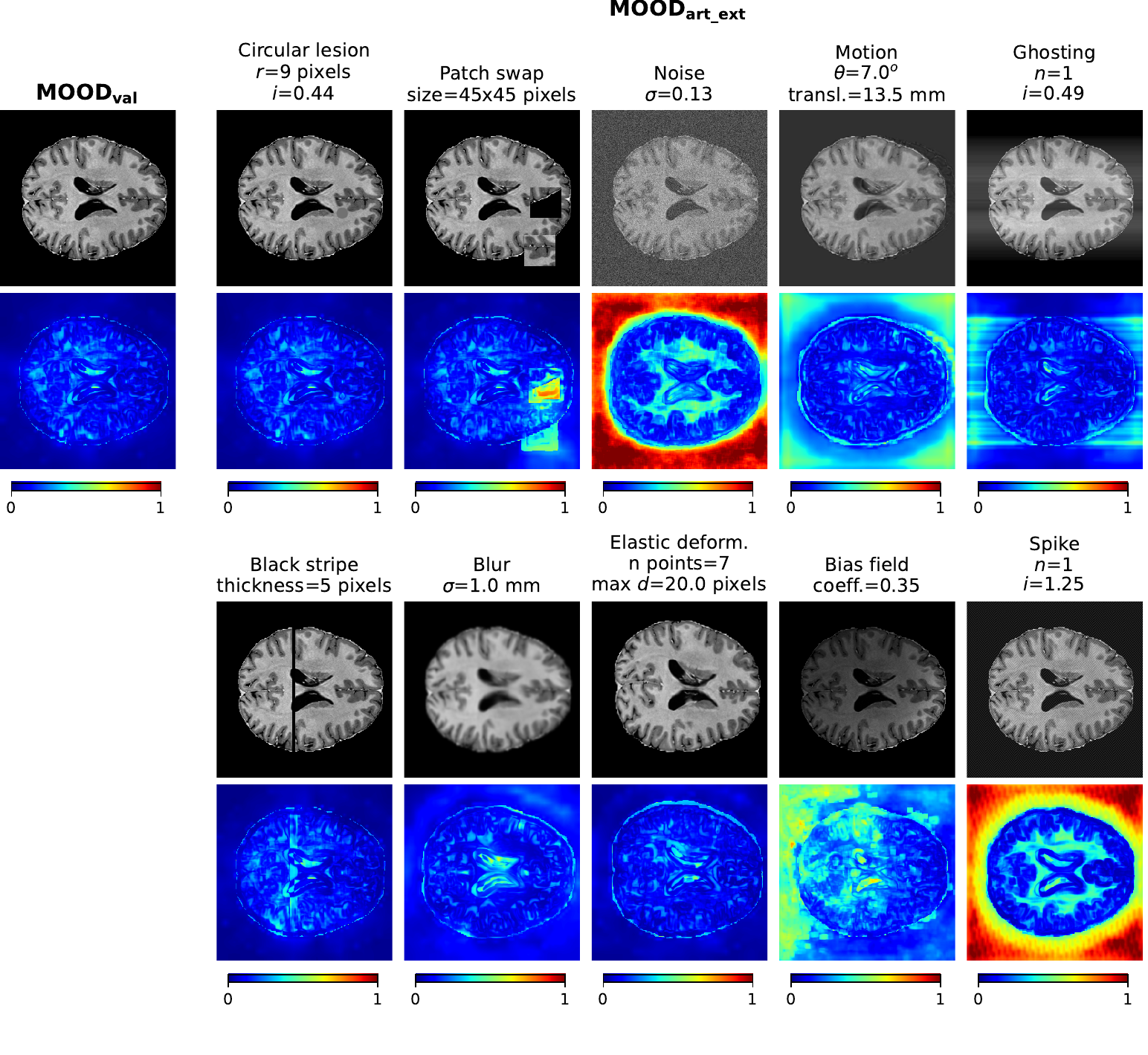}
    \caption{Example image from the MOOD\textsubscript{val} dataset (first row, first column) and the final prediction map $\mathcal{A}$ obtained from averaging LPIPS and $1-c(x,\hat{x})$ for both the $2\times2$ and $8\times8$ models (second row, first column). The following columns show the counterparts from MOOD\textsubscript{art\_ext} with their final prediction maps presented below. Additionally, the parameter values used to generate the artifacts are presented in accordance with the explanation provided in Table 1 in the main paper.}
    \label{fig:visual_results_transformed}
\end{figure*}

This subsection presents additional results for the extended artifact dataset. \autoref{fig:visual_results_transformed} shows an example image from MOOD\textsubscript{val} and its MOOD\textsubscript{art\_ext} counterparts from MOOD\textsubscript{art\_ext}, along with their corresponding final prediction maps $\mathcal{A}$. The visual comparison highlights the effect of the artifacts on the images and the resulting OOD detection predictions.

\begin{figure*}[h]
    \centering
    \includegraphics[width=\linewidth]{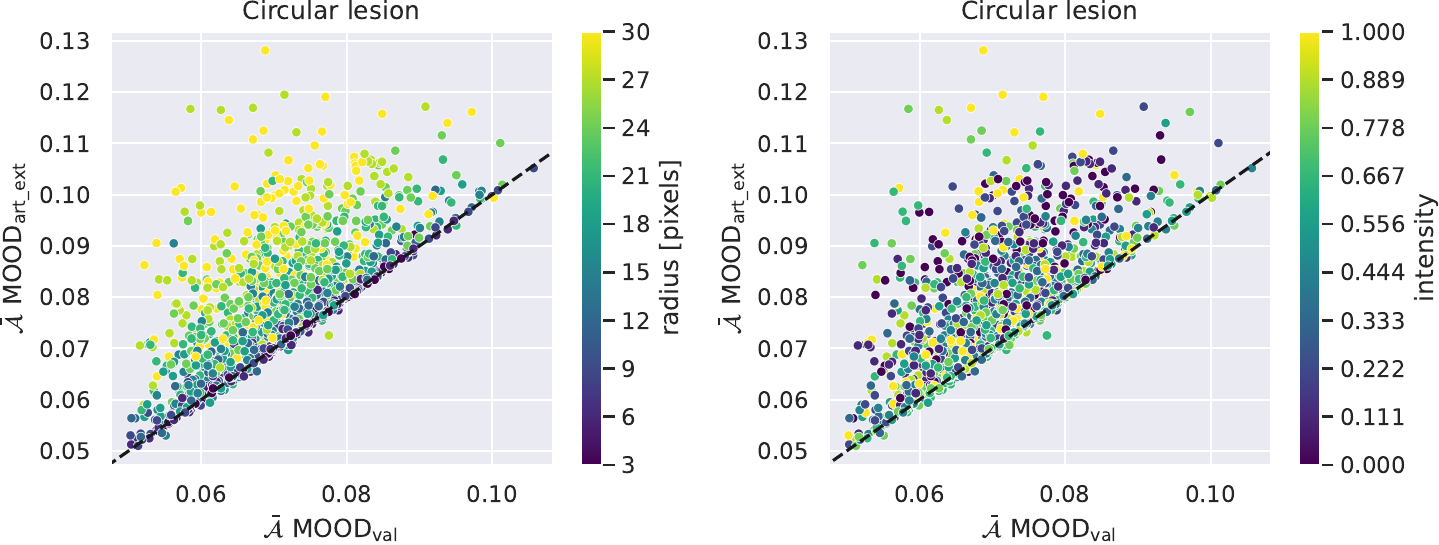}
    \caption{The score \(\bar{\mathcal{A}}\) for all slices with a circular lesion in MOOD\textsubscript{art\_ext} versus \(\bar{\mathcal{A}}\) of the same slice in the MOOD\textsubscript{val} dataset. The colors in the left and right figures indicate the different radii and intensities, respectively, used to the create the cases with circular lesions. The black dashed line indicates that \(\bar{\mathcal{A}}\) MOOD\textsubscript{art\_ext} is equal to \(\bar{\mathcal{A}}\) MOOD\textsubscript{val}.}
    \label{fig:corr_circlesion}
\end{figure*}

\begin{figure*}[h]
    \centering
    \includegraphics[width=\linewidth]{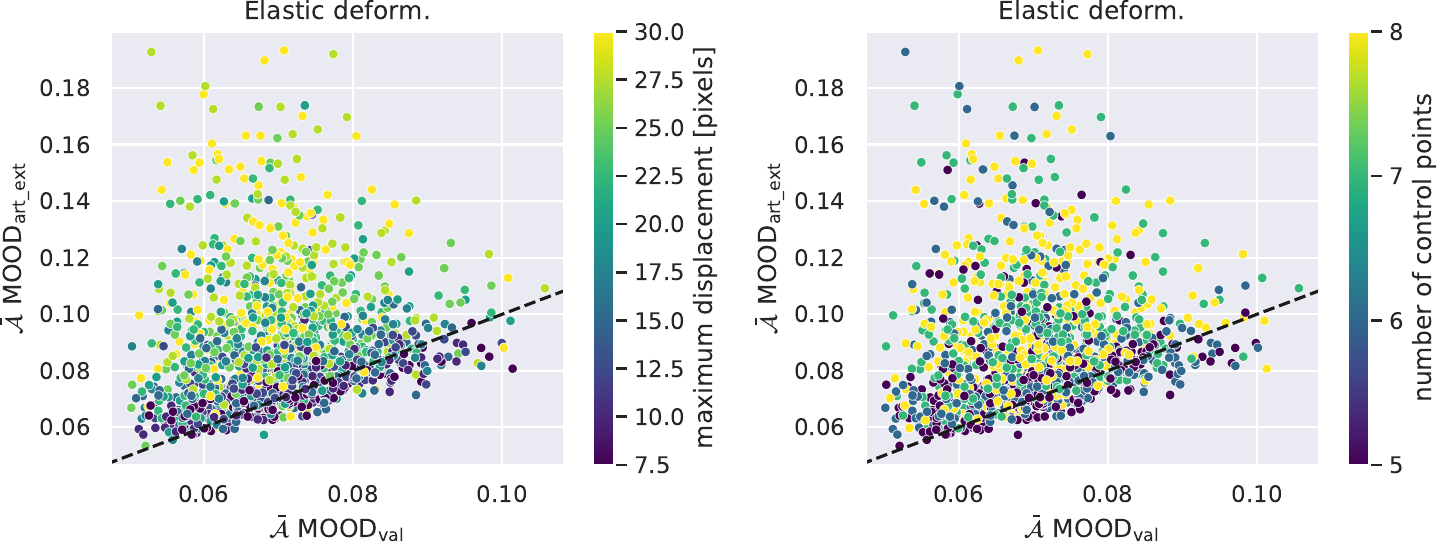}
    \caption{The score \(\bar{\mathcal{A}}\) for all deformed slices in MOOD\textsubscript{art\_ext} versus \(\bar{\mathcal{A}}\) of the same slice in the MOOD\textsubscript{val} dataset. The colors in the left and right figures indicate the different maximum displacements and number of control points, respectively, used to the create the cases with elastic deformation. The black dashed line indicates that \(\bar{\mathcal{A}}\) MOOD\textsubscript{art\_ext} is equal to \(\bar{\mathcal{A}}\) MOOD\textsubscript{val}.}
    \label{fig:corr_deform}
\end{figure*}

\begin{figure*}[h]
    \centering
    \includegraphics[width=\linewidth]{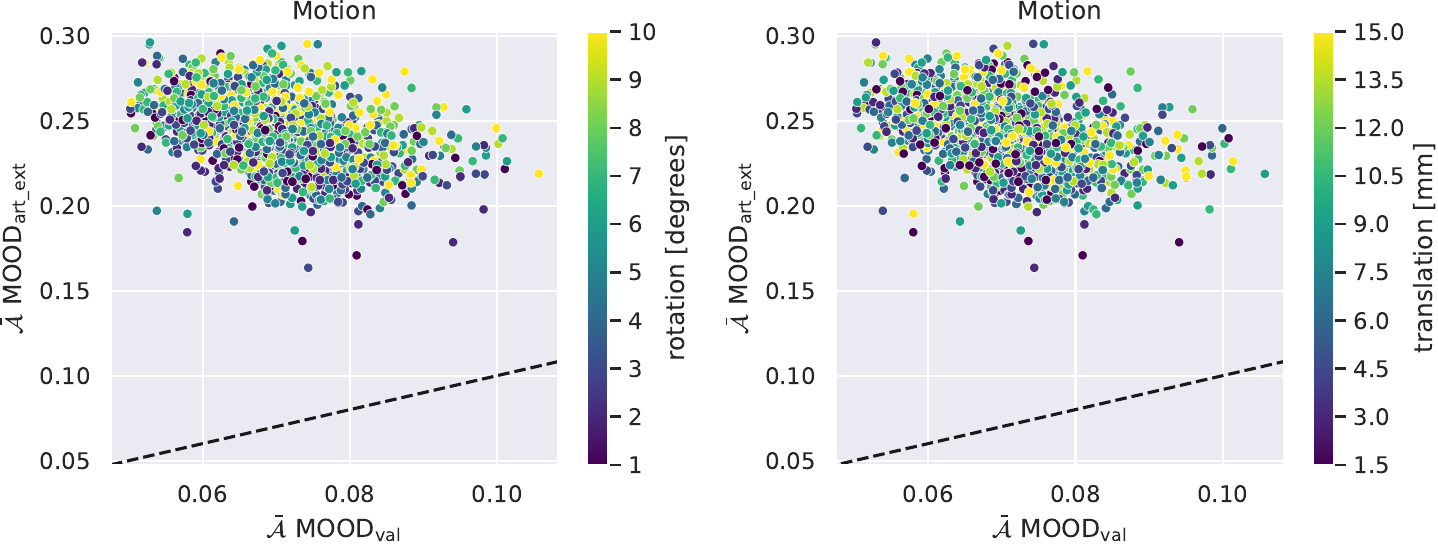}
    \caption{The score \(\bar{\mathcal{A}}\) for all slices with patient motion in MOOD\textsubscript{art\_ext} versus \(\bar{\mathcal{A}}\) of the same slice in the MOOD\textsubscript{val} dataset. The colors in the left and right figures indicate the different degrees of rotation and translation values, respectively, used to the create the cases with motion. The black dashed line indicates that \(\bar{\mathcal{A}}\) MOOD\textsubscript{art\_ext} is equal to \(\bar{\mathcal{A}}\) MOOD\textsubscript{val}.}
    \label{fig:corr_motion}
\end{figure*}

\begin{figure*}[h]
    \centering
    \includegraphics[width=\linewidth]{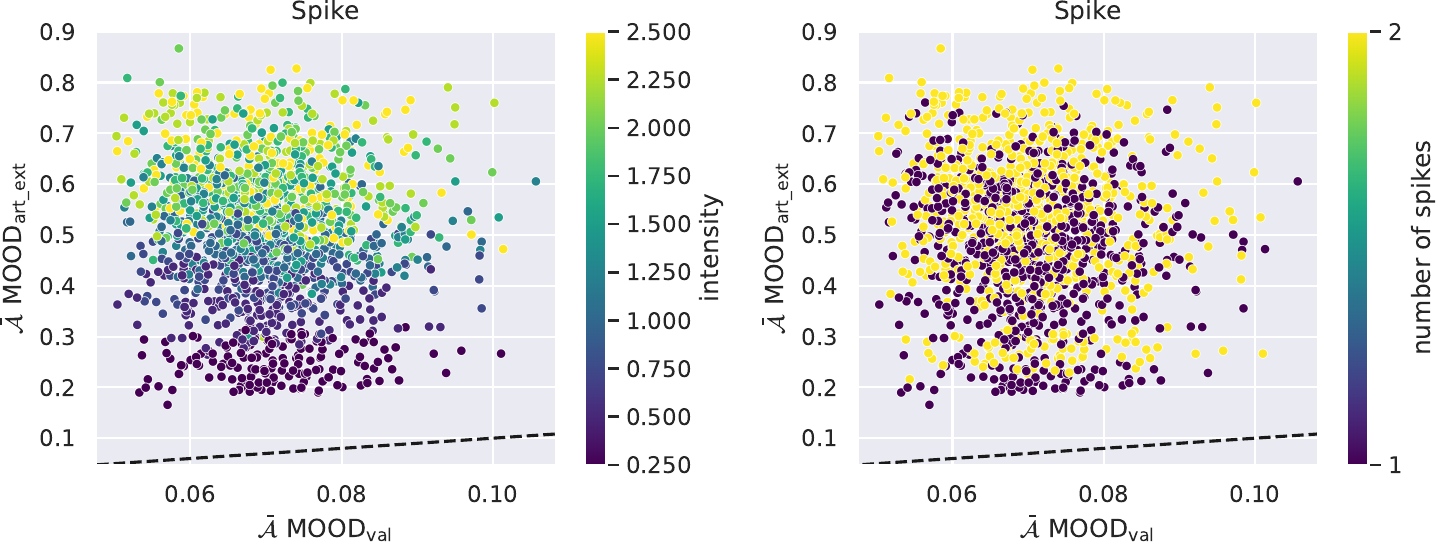}
    \caption{The score \(\bar{\mathcal{A}}\) for all slices with spikes in MOOD\textsubscript{art\_ext} versus \(\bar{\mathcal{A}}\) of the same slice in the MOOD\textsubscript{val} dataset. The colors in the left and right figures indicate the different intensities and number of spikes, respectively, used to the create the cases with spikes. The black dashed line indicates that \(\bar{\mathcal{A}}\) MOOD\textsubscript{art\_ext} is equal to \(\bar{\mathcal{A}}\) MOOD\textsubscript{val}.}
    \label{fig:corr_spike}
\end{figure*}

\begin{figure}[h]
    \centering
    \includegraphics[width=\linewidth]{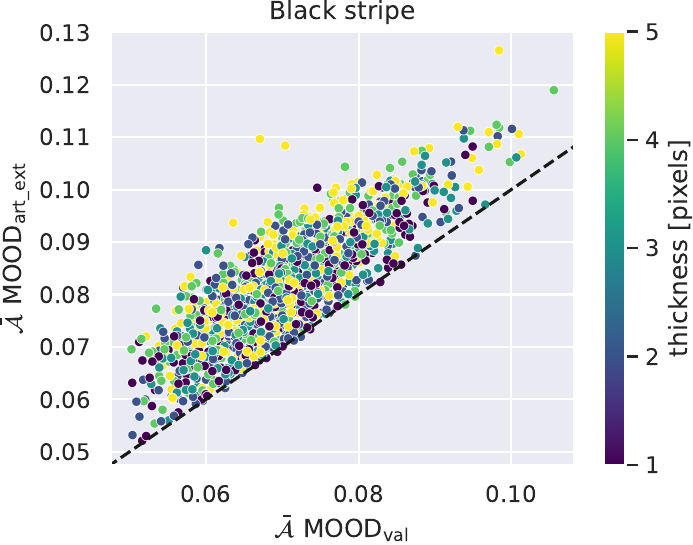}
    \caption{The score \(\bar{\mathcal{A}}\) for all black stripe slices in MOOD\textsubscript{art\_ext} versus \(\bar{\mathcal{A}}\) of the same slice in the MOOD\textsubscript{val} dataset. The colors indicate the different thicknesses used to the create the black stripe. The black dashed line indicates that \(\bar{\mathcal{A}}\) MOOD\textsubscript{art\_ext} is equal to \(\bar{\mathcal{A}}\) MOOD\textsubscript{val}.}
    \label{fig:corr_blackslice}
\end{figure}

\begin{figure}[h]
    \centering
    \includegraphics[width=\linewidth]{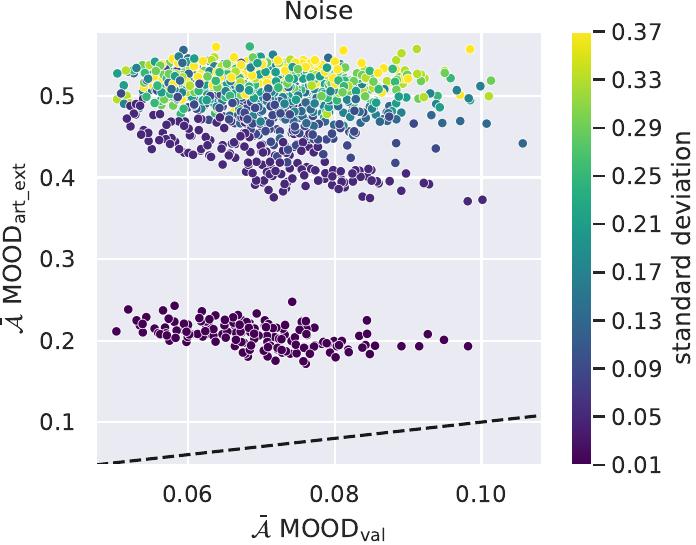}
    \caption{The score \(\bar{\mathcal{A}}\) for all noisy slices in MOOD\textsubscript{art\_ext} versus \(\bar{\mathcal{A}}\) of the same slice in the MOOD\textsubscript{val} dataset. The colors indicate the different standard deviations used to the create the noisy cases. The black dashed line indicates that \(\bar{\mathcal{A}}\) MOOD\textsubscript{art\_ext} is equal to \(\bar{\mathcal{A}}\) MOOD\textsubscript{val}.}
    \label{fig:corr_noise}
\end{figure}

\begin{figure}[h]
    \centering
    \includegraphics[width=\linewidth]{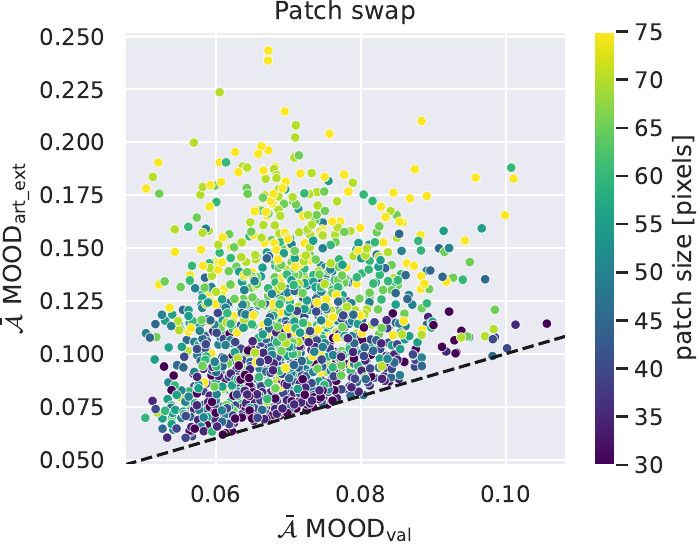}
    \caption{The score \(\bar{\mathcal{A}}\) for all slices with a patch swapped in MOOD\textsubscript{art\_ext} versus \(\bar{\mathcal{A}}\) of the same slice in the MOOD\textsubscript{val} dataset. The colors indicate the different patch sizes used to the create the patch swap cases. The black dashed line indicates that \(\bar{\mathcal{A}}\) MOOD\textsubscript{art\_ext} is equal to \(\bar{\mathcal{A}}\) MOOD\textsubscript{val}.}
    \label{fig:corr_swap}
\end{figure}

Additionally, for each artifact type, we present scatterplot(s) showing the correlation between the parameter value(s) and the score $\mathcal{\bar{A}}$. Three artifacts types already discussed in the main paper include bias field (Figure 12), blur (Figure 13), and ghosting (Figure 14). The other artifacts detailed in this appendix include circular lesions (\autoref{fig:corr_circlesion}), elastic deformation (\autoref{fig:corr_deform}), motion (\autoref{fig:corr_motion}), spike (\autoref{fig:corr_spike}), black stripe (\autoref{fig:corr_blackslice}), noise (\autoref{fig:corr_noise}), and patch swap (\autoref{fig:corr_swap}). These figures provide additional insight into how sensitive the proposed OOD detection model is to certain artifact parameters.

\twocolumn

